\documentclass[aps,physrev,reprint,groupedaddress]{revtex4-2}

\usepackage{graphicx}
\usepackage{dcolumn}
\usepackage{bm}
\usepackage{braket}
\usepackage{amsfonts}
\usepackage{hyperref}

\newcommand{\omegaa}{\omega_\mathrm{a}}

\begin{document}

\title{Collapse and Inversion of the Josephson Potential \\in a Strongly Driven Superconducting Circuit}

\author{Sercan Deve}
\email[These authors contributed equally; \\]{S.Deve@tudelft.nl, M.Villiers@tudelft.nl}
\affiliation{Kavli Institute of Nanoscience, Delft University of Technology, PO Box 5046, 2600 GA Delft, The Netherlands}

\author{Marius Villiers}
\email[These authors contributed equally; \\]{S.Deve@tudelft.nl, M.Villiers@tudelft.nl}
\affiliation{Kavli Institute of Nanoscience, Delft University of Technology, PO Box 5046, 2600 GA Delft, The Netherlands}

\author{Marijn A. Morssink}
\affiliation{Kavli Institute of Nanoscience, Delft University of Technology, PO Box 5046, 2600 GA Delft, The Netherlands}

\author{Robin C. Dekker}
\affiliation{Kavli Institute of Nanoscience, Delft University of Technology, PO Box 5046, 2600 GA Delft, The Netherlands}

\author{Gary A. Steele}
\email[]{G.A.Steele@tudelft.nl}
\affiliation{Kavli Institute of Nanoscience, Delft University of Technology, PO Box 5046, 2600 GA Delft, The Netherlands}

\date{September 28, 2026}

\begin{abstract}

Superconducting circuits embedding Josephson junctions leverage microwave drives for control and measurement of quantum states. Although increasing the drive power is desirable for improving the efficiency of these operations, it eventually triggers unwanted transitions to uncontrolled states. While careful choice of circuit symmetries and parameters can mitigate these effects, the presence of spurious circuit modes spoils the resilience to high power. In this work, we engineer a transmon-resonator system free of any detrimental unwanted transitions. This resilience enables us to access drive powers at which we uncover a remarkable physical phenomenon: the collapse and inversion of the Josephson potential. Through spectroscopy and readout experiments, we confirm that the driven potential goes to zero and inverts as the power increases. The inversion corresponds to the dynamical stabilization of the transmon at its unstable equilibrium point, directly analogous to an inverted pendulum. This result reveals a new limitation of strongly driven superconducting circuits beyond drive-induced transitions. In addition, the dynamical renormalization of the Josephson potential opens up novel avenues for the control of superconducting circuits, and the autonomous stabilization of noise-resilient quantum states.  

\end{abstract}

\maketitle

Superconducting circuits have benefited from the great versatility of microwave engineering to become a workhorse of quantum information processing \cite{Blais2021}. At their heart, Josephson junctions provide the nonlinearity necessary to perform nontrivial operations, while microwave drives activate it. The strength of these nonlinear interactions grows with the drive amplitude. As a consequence, stronger drives could accelerate qubit control \cite{Eickbusch2022, Valery2022} and readout \cite{Swiadek2024, Spring2025}, and improve coupler operation \cite{Caldwell2018,Lu2023}, the stabilization of noise-resilient quantum states \cite{Grimm2020, Lescanne2020}, or the Floquet engineering of Hamiltonians \cite{Gandon2022, Nguyen2024}. In the context of optomechanical systems interfaced with qubits, strong drives will be needed to prepare nonclassical mechanical states to test gravitational collapse models \cite{Gely2021}. Yet, the dynamic range of Josephson circuits is currently limited by drive-induced unwanted transitions to uncontrolled states \cite{Lescanne2019, Cohen2023, Xia2025, Dai2026}. One of the most prominent mechanisms responsible is the activation of multi-excitation resonances \cite{Sank2016,Dumas2024, Nesterov2024}, in which the junction nonlinearity mediates higher-order parametric mixing processes. These drastically limit the performance of Josephson circuits at high power.\\
\indent A solution to this problem is provided by careful circuit engineering. In coupler and mixing elements, leveraging circuit symmetries has been a standard approach for canceling spurious mixing processes and increasing dynamic range \cite{Roch2012,Lescanne2020,Grimm2020,Lu2023,Maiti2025}. Recently, the same approach has been applied to qubit readout \cite{Dassonneville2020,Salunkhe2025}. Here, it was shown that replacing the conventional linear dipole coupling between qubit and resonator by a native nonlinear coupling could eliminate all odd-parity transitions \cite{Chapple2025,Mori2025b,Hazra2025}. Other mitigation approaches include the cancellation of transitions through mixed coupling \cite{Beaulieu2026}, or asymptotic suppression by placing the resonator much higher in frequency than the qubit \cite{Kurilovich2025}. 
Nonetheless, the presence of spurious modes in the circuit or the environment can still significantly limit the maximum achievable drive power \cite{Dai2026,Hazra2026}.\\  
\indent In this work, we further extend the dynamic range of a Josephson circuit and reveal a novel physical phenomenon: the collapse and inversion of the Josephson potential. In a transmon-resonator system, we show that the instantaneous modulation of the Josephson potential imparted by a resonator drive can fully average it out, and subsequently revive it with opposite sign. Crucial to this observation is the implementation of a native cosine-cosine coupling free of activated multi-excitation resonances \cite{Hazra2026}. This collapse and inversion is first revealed in a spectroscopy experiment, which shows the qubit frequency going down to zero and recovering for increasing drive power. A subsequent readout experiment evidences the inversion through a $\pi$ phase shift of the localization of the transmon eigenstates, akin to the autonomous stabilization of a strongly driven pendulum in the upside-down position \cite{Stephenson1908,Butikov2001}. The collapse of the Josephson energy reveals a previously overlooked fundamental limitation on qubit readout and strongly driven Josephson circuits. Beyond strong driving limitations, the dynamical renormalization of the Josephson potential opens new opportunities in quantum information, such as the implementation of a fully protected $\cos 2\hat{\varphi}$ qubit through driving of a simple transmon circuit \cite{Venkatraman2022,Wang2024,Chowdhury2026}.

\section*{Concept} 
\begin{figure}[]
\includegraphics[width = 87mm]{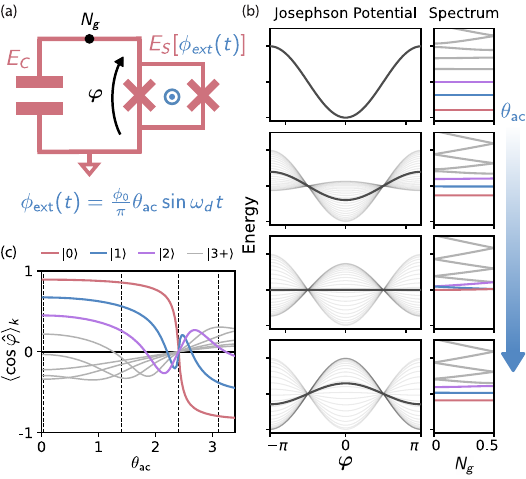}
\caption{\label{fig:fig1} \textbf{Concept.} (a)~A transmon circuit with charging energy $E_C$ embedding a symmetric SQUID with Josephson energy $E_S[\phi_\mathrm{ext}(t)]$ is flux-driven at frequency $\omega_d$ with a reduced amplitude $\theta_\mathrm{ac}$. (b)~Potential energy vs phase $\varphi$ (left) and spectrum vs offset charge $N_g$ (right) under increasing drive amplitude (top to bottom, amplitudes marked as vertical dashed lines in panel (c)). The thin potential lines are instantaneous profiles plotted during one drive period, while the thick ones are time-averaged. The spectra are computed with the averaged potentials. 
(c)~Diagonal matrix element of $\cos\hat{\varphi}$ for the eigenstates $\ket{k}$ of Hamiltonian Eq.~(\ref{eq:Heff}) and averaged over offset charge, vs reduced drive strength.}
\end{figure}

The flux-tunable transmon is a prototypical superconducting circuit \cite{Koch2007,Blais2021}. It consists of an island with charging energy $E_C$, shunted to ground by a SQUID with Josephson energy $E_S(\phi_\mathrm{ext})$ where $\phi_\mathrm{ext}$ is the magnetic flux threading its loop  (Fig.~\ref{fig:fig1}(a)). For a SQUID made out of identical tunnel junctions with Josephson energy $E_J/2$, one finds: $E_S(\phi_\mathrm{ext})=E_J\cos\pi\phi_\mathrm{ext}/\phi_0$, where $\phi_0$ is the flux quantum. This circuit supports one mode described in terms of the excess number of Cooper pairs on the island $\hat{N}$, and its conjugate variable $\hat{\varphi}$, the superconducting phase drop across the SQUID. The transmon regime requires $|E_S(\phi_\mathrm{ext})|/E_C \gg 1$, for which the circuit's lowest eigenstates are bound states of the Josephson potential $-E_S(\phi_\mathrm{ext})\cos\hat{\varphi}$ (Fig.~\ref{fig:fig1}(b) top). Their localization in phase and converse delocalization in Cooper-pair number grants them a reduced sensitivity to fluctuations of the offset charge $N_g$, a highly desirable feature to encode quantum information in solid-state systems.\\ 
\indent A time-dependent flux drive $\phi_\mathrm{ext}(t) = \phi_\mathrm{ac}\sin\omega_dt$ modulates the Josephson energy and eventually changes its sign over time (thin potential lines in Fig.~\ref{fig:fig1}(b)). If the drive frequency $\omega_d$ is higher than the transmon transition frequencies, this modulation can be averaged out (thick lines) (Supplementary Information). The driven circuit is thus described by the renormalized transmon Hamiltonian:
\begin{equation}
    \hat{\mathcal{H}} = 4E_C(\hat{N}-N_g)^2 - E_JJ_0(\theta_\mathrm{ac}) \cos\hat{\varphi}\;,
    \label{eq:Heff}
\end{equation}
where $J_0$ is the zeroth-order Bessel function of the first kind, and $\theta_\mathrm{ac}=\pi\phi_\mathrm{ac}/\phi_0$ is the reduced flux-drive amplitude. Crucially, the Bessel function oscillates with a first zero at $\theta_\mathrm{ac}\approx2.405$, and a slowly decaying envelope $\sim 1/\sqrt{\theta_\mathrm{ac}}$. As a consequence, a high-frequency flux drive can lower the height of the average Josephson potential, until it collapses, and revives in an inverted fashion. \\
\indent This sequence has clear spectroscopic features (Fig.~\ref{fig:fig1}(b)). First, as the drive amplitude increases, the circuit transition frequencies are reduced and charge sensitivity is restored, bringing the system towards a dynamical Cooper-pair box regime. Then, at the collapse point, the tunneling of Cooper pairs originally mediated by the junctions is destroyed, and the spectrum becomes gapless. We note that a similar averaging has been analyzed in the context of a strongly driven fluxonium qubit \cite{Lewellen2026}. Finally, in the inverted regime, finite transition frequencies and charge insensitivity recover, though never fully due to the decay of $J_0(\theta_\mathrm{ac})$.\\
\indent The revival of the Josephson potential in the inverted regime corresponds to a dynamical stabilization around $\varphi=\pi$. In Fig.~\ref{fig:fig1}(c), we show the expectation values of $\cos\hat{\varphi}$ for the eigenstates of Hamiltonian Eq.~(\ref{eq:Heff}). Their sign change at the collapse point is a manifestation of a $\pi$-shift of the phase localization (Supplementary Information). We shed light on this result using the transmon analogy with a rigid pendulum \cite{Koch2007}, where the phase plays the role of the angular position, the Josephson potential maps to the gravitational one, and the flux drive amounts to a modulation of gravity. In practice, gravity is effectively modulated in the inertial frame of a pendulum whose anchor point is shaken up and down. There, it has been long known that fast and strong modulation can stabilize upside-down motion, that is $\pi$-shifted from the rest position \cite{Stephenson1908}. In the following, we present an experiment which, to the best of our knowledge, gives the first evidence of this phenomenon in a superconducting circuit.

\section*{Implementation}
\begin{figure}[t]
\includegraphics[width = 87mm]{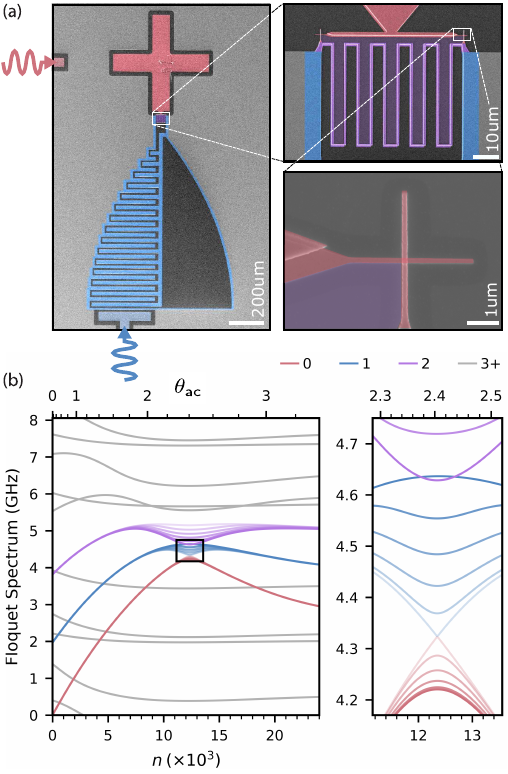}
\caption{\label{fig:fig2} \textbf{Implementation.} (a)~False-colored SEM picture of the device, with the transmon island ($E_C/h=103$~MHz) and the SQUID junctions ($E_J/2h=2.6$~GHz) in salmon, and the readout resonator in blue. Drive ports and input signals (wiggly arrows) have matching colors. A meander inductor (purple) forms the lower arm of the SQUID (area in light purple). Insets: zoom-in on the SQUID coupler and one junction. (b)~Floquet spectrum (lines, first 12 levels) of the transmon flux-driven at $\omegaa/2\pi=8.054$~GHz vs resonator occupation. Levels 0 to 2 are plotted with varying offset charge from 0 to 0.5 (dark to light colors). Levels above 3 are plotted with $N_g=0.2$. Inset: zoom-in on the collapse region.}
\end{figure}

Though introduced for the specific case of a flux-driven transmon embedding a symmetric SQUID, the collapse and inversion of the Josephson potential is a far more general feature of Josephson circuits under strong drives (Supplementary Information). We believe it has remained inaccessible up to now due to the susceptibility of existing circuits to multi-excitation resonances \cite{Tuorila2010,Wu2019,Mori2025b,Hazra2026}. These unwanted transitions to non-computational states -- either intrinsic to the transmon spectrum or mediated by spurious degrees of freedom -- have been extensively studied in the context of qubit readout \cite{Fechant2025,Wang2026,Dai2026, Hazra2026}. \\
\indent Here we present such an experiment, where a grounded transmon embedding a symmetric SQUID is read out through a longitudinally coupled resonator (Fig.~\ref{fig:fig2}(a)). The latter, with frequency $\omegaa/2\pi=8.054$~GHz and linewidth $\kappa/2\pi=3.1$~MHz, is a quarter wavelength coplanar-waveguide shunted to ground in a hook below the transmon. This unconventional geometry is designed to push spurious modes up in frequency, while maximizing the fraction $p_I=0.49$ of the resonator current flowing through the lower arm of the SQUID (Supplementary Information). This meander contributes to a fraction $p_L=0.48$ of the resonator inductance, and induces the native cosine-cosine coupling that powers the longitudinal interaction \cite{Potts2025, Salunkhe2025,Dassonneville2020, Mori2025b, Hazra2025, Chapple2025, Hazra2026}. Populating $n$ photons in the resonator effectively flux-drives the transmon with $\theta_\mathrm{ac} = p \varphi_{zpf} \sqrt{n}$, where $p=p_Ip_L$ is the coupling dilution factor, and $\varphi_{zpf}=0.092$ are the zero-point phase fluctuations of the resonator. At the sweet spot, it translates into a non-perturbative cross-Kerr interaction with amplitude $\chi_z/2\pi=-122$~kHz in the small $n$ limit. Finally, the transmon qubit frequency $\omega_{01}/2\pi=1.972$~GHz places the system in the regime $\omegaa/\omega_{01}\gg1$. Together with the symmetry of the cosine-cosine coupling, it is key to limiting the impact of multi-excitation resonances \cite{Kurilovich2025}. \\
\indent We study their prevalence in our design using the Floquet method, under a semiclassical approximation that treats the resonator in the correspondence limit \cite{Grifoni1998,Dumas2024}. It relies on the analysis of the Floquet spectrum of the transmon states dressed by the drive (Fig.~\ref{fig:fig2}(b)). Computed numerically for a drive at the resonator frequency, the energy levels are defined modulo $\omegaa$. Thus, a single Brillouin zone $[0, \omegaa]$ is sufficient to characterize the spectrum, where resonances between two levels can be attributed to high-order parametric processes mixing an integer number of drive photons. Those infamous multi-excitation resonances are scarce in the computed spectrum, and the remaining ones are either canceled by the coupling symmetry, or heavily suppressed in our parameter regime (Supplementary Information). Around $12.3\times10^3$ photons, a unique feature appears. First, the dressed bound states restore their charge sensitivity. Then, as the flux-drive amplitude coincides with $\theta_\mathrm{ac}\approx2.405$, the dressed levels 0 and 1 are either degenerate when $N_g=0.5$, or spaced by $\sim4E_C$ when $N_g=0$. A similar feature would be expected for a charged island in the absence of tunneling, and we interpret it as the dynamical cancellation of the Josephson energy. Remarkably, this spectrum is almost entirely captured by the averaged model, if not for small deviations around the collapse point arising from a higher-order $\cos2\hat{\varphi}$ correction (Supplementary Information).\\
\indent Fortified by the absence of spurious resonances in the system, we extend the averaged theory to incorporate the resonator response. In the presence of a strong resonant drive, the native cross-Kerr interaction can be linearized \cite{Touzard2019, Chapple2025}. It yields a displacement of the resonator field conditioned on the transmon state, with amplitude:
\begin{equation}
    \delta\alpha_{\ket{k}}(n,N_g) = p\varphi_{zpf} \frac{E_J}{\hbar\kappa} J_1(\theta_\mathrm{ac}) \times
    \langle \cos\hat{\varphi} \rangle_k (n,N_g)\:,
    \label{eq:chi_i}
\end{equation}
where $J_1$ is the first-order Bessel function of the first kind, and $\langle \cos\hat{\varphi} \rangle_k (n,N_g)$ is the $k^\mathrm{th}$ expectation value computed in the Floquet picture. Crucially, these values follow the same trajectory as the averaged model (Fig.~\ref{fig:fig1}(c)) (Supplementary Information). Overall, the native cosine-cosine coupling offers a unique possibility to image the dynamical phase shift of the transmon localization in a readout experiment.
 
\section*{Spectroscopic observation of collapse and inversion}  

\begin{figure}
\includegraphics{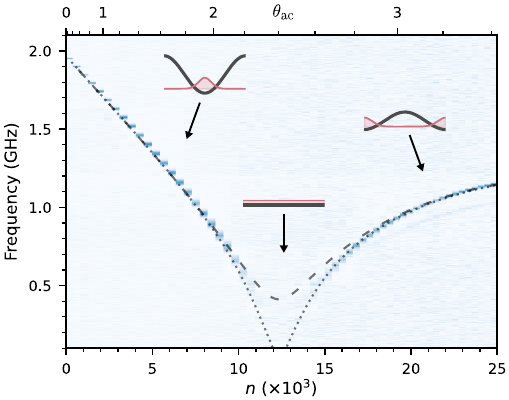}
\caption{\label{fig:fig3} \textbf{Qubit spectroscopy through collapse and inversion.}
Frequency shift of the qubit transition measured by two-tone spectroscopy as a function of resonator occupation. The color represents the digitizer signal in arbitrary units. The black dashed and dotted lines are fits to the measured spectrum based on diagonalization of the Hamiltonian Eq.~(\ref{eq:Heff}) for offset charges $N_g = 0.0$ and $N_g = 0.5$, respectively. Above, we indicate the renormalized Josephson potential profiles in black, and the corresponding ground-state wavefunctions in phase representation in salmon. The resonator occupation is calibrated from the fit. The top axis indicates the equivalent flux-drive amplitude $\theta_{\mathrm{ac}}$.} 
\end{figure}

As a first probe into the dynamical renormalization of the Josephson potential, we track the qubit frequency as a function of resonator drive power. The frequency is determined by performing a two-tone spectroscopy measurement, shown in Fig.~\ref{fig:fig3}. We identify three distinct regimes in the evolution of the qubit transition. First, the frequency decreases with increasing power, while simultaneously its linewidth broadens. Next, around $12.3\times10^3$ photons, the signal vanishes completely. Finally, we observe its revival, followed by an increase of the qubit frequency and narrowing of the linewidth. This nontrivial power dependence excludes typical effects such as power broadening and heating. In addition, the ability to resolve the transition over such a wide range of powers demonstrates the robustness of our circuit to multi-excitation resonances.\\
\indent We fit this nonlinear qubit frequency shift by diagonalizing the renormalized Hamiltonian of Eq.~(\ref{eq:Heff}) for offset charges $N_g = 0$ and $N_g = 0.5$. The model contains a single fit parameter used for converting the resonator input power to photon number. Prior to this simple conversion, a nonlinear rescaling of the input power was required to account for a nonmonotonic frequency shift of the resonator and the finite dynamic range of our room-temperature microwave components (Supplementary Information). The fit excellently captures the evolution of the qubit frequency, allowing us to interpret the three regimes in power as the flattening, collapse and inversion of the Josephson potential. This interpretation is strengthened by the observation of two intriguing features. We begin with the vanishing of the signal around $12.3 \times 10^3$ photons. Coinciding with the separation of the fitted lines, we attribute it to a reduction of the system coherence as charge sensitivity is restored. Next, the faint signature at high power is identified to be the 1-2~transition, negatively detuned from the qubit transition by the anharmonicity $\alpha/2\pi$. While in the inverted regime we measure $\alpha/2\pi \approx-E_C/h$, near the collapse $|\alpha|$ goes down, as expected when going from the transmon to the Cooper-pair box limit (Supplementary Information).\\
\indent Overall, the strong agreement with our simple model validates the dynamical renormalization of the Josephson potential. In Fig.~\ref{fig:fig3}, we plot its profile for the three aforementioned regimes, together with the ground-state wavefunctions in phase representation. They illustrate the delocalization at the collapse point, and the relocalization around $\varphi=\pi$ in the inverted regime. However, since qubit spectroscopy is only sensitive to the absolute value of the Josephson energy, one could argue that this measurement does not provide direct evidence of the inverted regime. In the next section, we construct a measurement that allows us to unambiguously reveal both the collapse and inversion of the Josephson potential.

\section*{Observation of collapse and inversion through readout}

\begin{figure}
\includegraphics{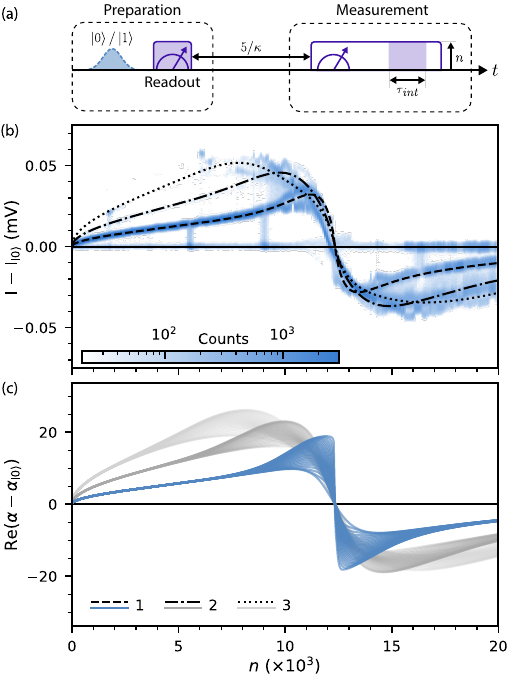}
\caption{\label{fig:fig4} \textbf{Collapse and inversion revealed through readout} (a)~Pulse sequence. An optional $\pi$ pulse prepares the transmon in either the ground $\left | 0 \right \rangle$ or first excited $\left | 1 \right \rangle$ state, followed by a post-selection readout pulse. After an idling time of $5/\kappa$, a long measurement pulse is applied. IQ distributions are acquired by integrating over an interval of length $\tau_{int} = 600$~ns at the end of the pulse (shaded area), ensuring a measurement of the steady-state resonator field only. We repeat this sequence for increasing measurement amplitude, which we convert into a resonator steady-state occupation $n$. 
(b)~Measured histogram of the I quadrature for transmon prepared in $\left | 1 \right \rangle$ as a function of steady-state resonator occupation. At each $n$, the quadrature is referenced to the one of the dressed ground state $I_{|0\rangle}$ determined from the IQ distribution for preparation in $\ket{0}$ (Supplementary Information). The color indicates the number of counts. A clear sign reversal of the readout signal is observed close to $12.3\times10^3$ photons. (c)~Semiclassical simulation of the measurement pulse for different initial transmon states. At each $n$, the steady-state intracavity field $\alpha$ is referenced to the one of the dressed ground state $\alpha_{|0\rangle}$. Each trajectory is plotted for $N_g\in[0,0.5]$. In (b), the offset-charge averaged resonator fields are plotted in dashed lines. Scaling of the y-axis is the only free parameter used to match simulation to experimental data.}
\end{figure}

To confirm the inversion of the potential, we make the most of the cosine-cosine coupling by probing the expectation values of $\cos\hat{\varphi}$ through a readout experiment \cite{Blais2021}. There, information about the transmon state is encoded in the conditional displacement of Eq.~(\ref{eq:chi_i}), which we measure by reflecting a near-resonant microwave drive from the resonator. In Fig.~\ref{fig:fig4}(a), we present such an experiment. After preparing the transmon in either the ground $\ket{0}$ or first excited $\ket{1}$ state, a measurement pulse of varying amplitude is applied. Since this pulse is also responsible for dressing the transmon, integration is only performed at the end to ensure probing the dressed states with the resonator in steady state.

In Fig.~\ref{fig:fig4}(b), we plot the histogram of the I quadrature for preparing $\left | 1 \right \rangle$ against steady-state resonator occupation $n$. The separation between the dressed excited and ground states grows with increasing resonator occupation, which can be understood as an improvement of the signal-to-noise ratio. At specific photon numbers (e.g. around $1.5\times10^3$, $5.5\times10^3$, and $9\times10^3$), population is transferred to other states, as a consequence of drive-induced transitions involving spurious extrinsic degrees of freedom, such as two-level systems \cite{Dai2026}. From $n>8\times10^3$, we observe an increase in the population of higher excited states, which can be well understood through the dynamical collapse of the potential (Supplementary Information). Finally, around $12.3\times10^3$ photons, the readout signal for all observed states suddenly crosses zero and subsequently recovers with opposite sign. 

We perform a semiclassical simulation of the measurement pulse for different initial transmon states (see Supplementary Information for simulation details), and plot the steady-state intracavity field in Fig.~\ref{fig:fig4}(c). As highlighted in Fig.~\ref{fig:fig4}(b) by overlaying the offset-charge averaged trajectories, there is a clear agreement between simulation and data. It finds its origin in the expectation values of $\cos \hat{\varphi}$. They are computed using Floquet analysis and show the same evolution as the ones predicted by the averaged theory (Fig.~\ref{fig:fig1}(c)). Specifically, they all collapse around $12.3\times10^3$ photons and revive with opposite sign. The complete inversion of the readout signal, proportional to $\cos\hat{\varphi}$ according to Eq.~(\ref{eq:chi_i}), is a key signature of the localization of the transmon in the $\pi$-phase state, providing unequivocal evidence of the dynamical stabilization of the circuit in the inverted pendulum position. 

\section*{Conclusion and Outlook} 

In summary, we have implemented a transmon-resonator circuit resilient to multi-excitation resonances, through the combination of a symmetry-protected coupling and large detuning. Its exceptional resilience has allowed us to observe the collapse and inversion of the Josephson potential, a result of its high-frequency modulation imparted by a strong resonator drive. Through spectroscopy, we measure a nonlinear qubit frequency shift consistent with the renormalization of the potential. A readout experiment reveals a $\pi$-shift of the localization of the transmon states after the collapse. Contrary to tuning the static flux which shifts the equilibrium position, 
the high-frequency flux drive dynamically stabilizes the unstable equilibrium point, a direct quantum analog of an inverted classical pendulum. This stabilization mechanism needs however to be distinguished from the Kapitza regime, which relies on the higher-order $\cos2\hat{\varphi}$ correction to the averaged potential \cite{Butikov2001}. \\
\indent The collapse and inversion of the Josephson potential sets a new limitation for strongly driven Josephson circuits. Since the strength of drive-activated processes depends on the Josephson energy, careful experimental design is required to take advantage of its dynamical renormalization during high-power operation. For the case of the transmon, we highlight several practical implications. First, as the Josephson energy decreases, charge sensitivity is restored, exposing the system to charge-induced decoherence. Next, as the spectrum becomes gapless at the collapse point, driving through this region triggers unwanted transitions through Landau-Zener processes \cite{Grifoni1998, Ivakhnenko2023}. Finally, the transmon is expected to suffer from measurement-induced dephasing due to the resonant nature of the resonator drive.\\
\indent Fortunately, standard control techniques can alleviate these limitations. Detuning the drive is expected to significantly reduce the measurement-induced dephasing rate \cite{Lescanne2019}. Control of the offset charge in combination with pulse shaping will mitigate the charge sensitivity and Landau-Zener transitions near the collapse point. This will enable the quantum control of the transmon near the collapse and in the inverted regime.\\
\indent The robustness of our circuit to multi-excitation resonances lets us engineer its Josephson harmonics through driving. In recent years, circuits featuring high-order harmonics have received great interest for the stabilization of noise-resilient quantum states, of which the 0-$\pi$ qubit is a prime example \cite{Brooks2013}. They typically feature mesoscopic junctions \cite{Larsen2020, Brosco2024, Feldstein-Bofill2026}, or multiple circuit-modes \cite{Roverch2026,Shagalov2026,Zhurbina2026}, both sharing the additional challenge of canceling the bare $\cos \hat{\varphi}$ contribution. In contrast, our circuit uniquely produces an effective $\cos 2\hat{\varphi}$ contribution at higher order of the potential averaging \cite{Venkatraman2022,Wang2024,Chowdhury2026}, while simultaneously canceling the $\cos \hat{\varphi}$ contribution at the collapse point. This paves the way for a fully dynamical implementation of an intrinsically protected qubit.

\begin{acknowledgments}
M.V. acknowledges a fruitful discussion with Samuel Cailleaux on inverted pendulums. This publication is part of the project ‘Superconducting Electromechanics: Massive superpositions for exploring quantum mechanics and general relativity.’, project number VI.C.212.087 of the research programme VICI round 2021, financed by the Dutch Research Council (NWO). R.C.D. acknowledges support from the Netherlands Organisation for Scientific Research (NWO/OCW) as part of the Frontiers of Nanoscience program.

\textbf{Author contributions:} S.D. contributed to conceptual development, design, simulations, theory, fridge operation and writing of the manuscript, and led fabrication, measurement and data analysis. M.V.  co-supervised all aspects of the project, and contributed to conceptual development, design, measurement, simulations, theory, fridge operation, and writing of the manuscript. M.A.M. developed initial theory and data analysis of collapse and inversion, helped develop measurement protocols, provided insight into device parameters required for observing collapse and inversion, and provided theory input for the manuscript. R.C.D. built and characterized the parametric amplifier, contributed to the design, fabrication, and development of earlier generations of the device, and helped establish the experimental platform that enabled this work. G.A.S. supervised the project, provided the lab infrastructure, conceptualized and gave feedback on the figures and storyline, contributed to the understanding of the inverted pendulum regime, and gave feedback on the text of the manuscript.

\textbf{Competing interests:} The authors declare no competing interests.

\textbf{Data and materials availability:} All data, analysis code, and measurement software will be made available in a Zenodo repository.

\end{acknowledgments}

\bibliography{bibliography.bib}

\end{document}


\addtocontents{toc}{\protect\setcounter{tocdepth}{0}}

\clearpage
\onecolumngrid

\setcounter{equation}{0}
\setcounter{figure}{0}
\setcounter{table}{0}
\renewcommand{\theequation}{S\arabic{equation}}
\renewcommand{\thefigure}{S\arabic{figure}}

\noindent\textbf{\textsf{\Large Supplementary Information: Collapse and Inversion of the Josephson Potential in a Strongly Driven Superconducting Circuit}}

\vspace{2mm}

\normalsize
\vspace{.3cm}

\noindent\textsf{Sercan Deve,$^*$ Marius Villiers,$^*$ Marijn A. Morssink, Robin C. Dekker, and Gary A. Steele$^\dagger$}

\vspace{.2cm}
\noindent\textit{Kavli Institute of Nanoscience, Delft University of Technology, PO Box 5046, 2600 GA Delft, The Netherlands}\\

\firstpagefootnoteone{$^*$ These authors contributed equally; S.Deve@tudelft.nl, M.Villiers@tudelft.nl}
\firstpagefootnotetwo{$^\dagger$ G.A.Steele@tudelft.nl}

\tableofcontents

\clearpage

\section{Experimental Setup}

\subsection{Fabrication}
The device is fabricated on a 10$\times$10~mm$^2$ high-resistivity silicon chip using a single lithography-step aluminum lift-off process. The fabrication begins by spinning a bi-layer resist stack (320 nm MAA EL8 and 720 nm PMMA 950k A6) and patterning the full design, i.e. bulk circuitry and junctions, using electron-beam lithography. The bi-layer resist is developed in a cold 1:3 H$_2$O:IPA solution for two minutes, followed by a 15-second rinse in IPA. Prior to metal deposition, the surface is treated by an oxygen plasma descum (100 W, 200 sccm, 1 minute) and a clean in BOE 7:1 for one minute. Double-angle shadow evaporation of aluminum with an intermediate in-situ oxidation is performed in a Plassys MEB550. This process forms our Manhattan-style Josephson junctions as well as all surrounding circuit elements and the ground plane. The bottom and top aluminum layers are 25~nm and 100~nm thick, respectively, evaporated at a rate of 1 nm/s and at an angle of 35$^{\circ}$ with respect to the surface normal of the chip. In between aluminum evaporations, oxidation is carried out at a pressure of 10~mbar for a duration of 10 minutes. At the end of the process, the aluminum is capped by a 10-minute oxidation at 1.3~mbar. The fabrication is completed by lift-off in NMP at 80~$^{\circ}$C and additional cleaning in acetone and IPA.

\subsection{Device Parameters}

\begin{table}[h!]
\begin{center}
\begin{tabular}{llc}
\hline \hline
Parameter name  & Symbol           & Value  \\ \hline
Resonator frequency & $\omegaa/2\pi$~(GHz) & 8.0536 \\
Resonator linewidth & $\kappa /2\pi$~(MHz) &  3.1\\
Transmon qubit frequency & $\omega_{01}/2\pi$~(GHz)  & 1.9716 \\
Transmon charging energy & $E_C/h$ (MHz) &  103\\
Josephson energy & $E_J/2h$~(GHz) &  2.61 \\
Junction asymmetry & $d$ & $< 0.03$ \\
Non-perturbative cross-Kerr & $\chi_z/2\pi$~(kHz) & -122 \\
Relaxation time & $T_1$~($\mu$s) &  71\\
Ramsey dephasing time & $T_2^R$~($\mu$s) & 56 \\
Hahn-echo dephasing time & $T_2^E$~($\mu$s) & 61 \\ 
Transmon effective temperature & $T_{\mathrm{eff}}$~(mK) & 35\\
\hline \hline
\end{tabular}
\label{tab:Device_params}
\caption{Measured device parameters. Average values are reported for $T_1$, $T_2^R$, $T_2^E$ and $T_{\mathrm{eff}}$.}
\end{center}
\end{table}

\subsection{Wiring Diagram}
\begin{figure}[]
\includegraphics{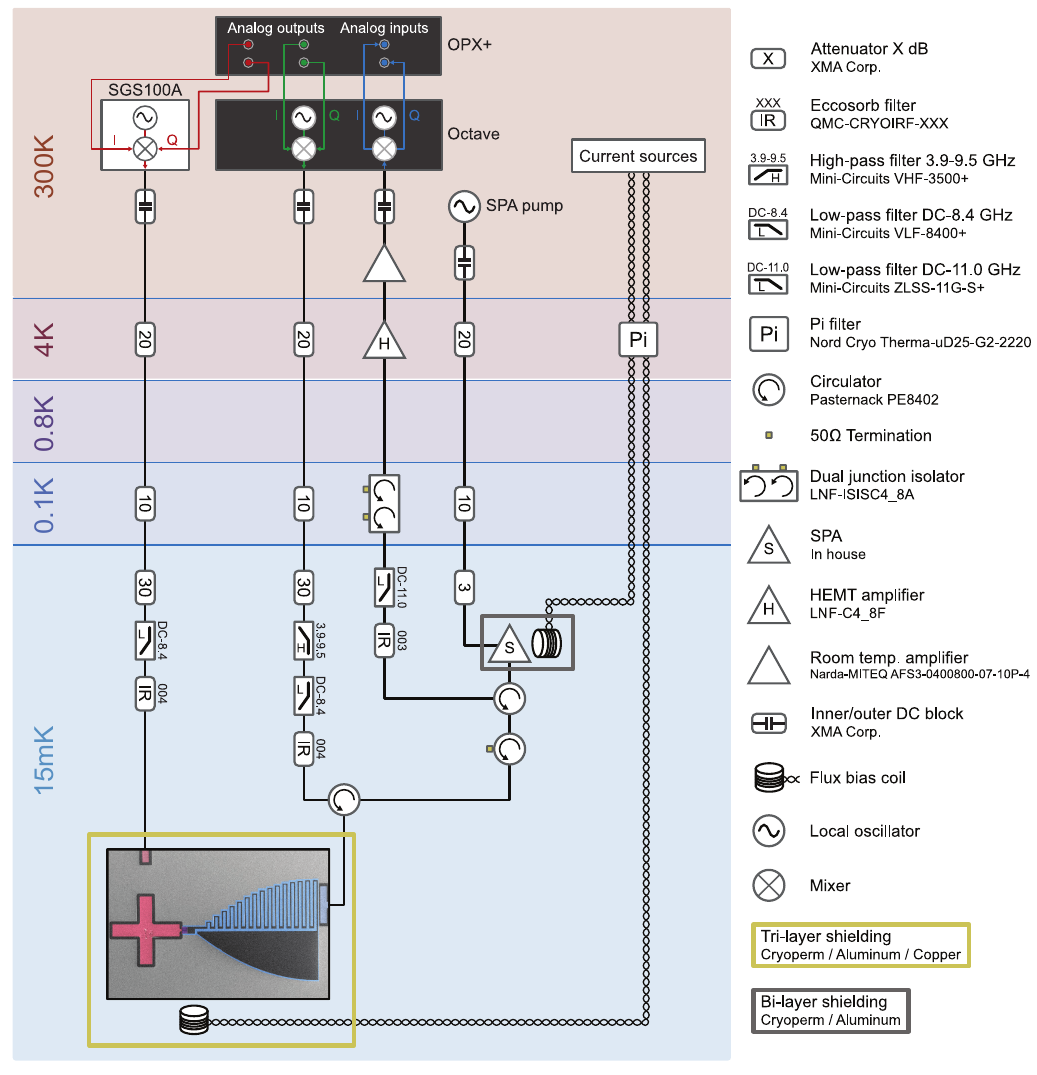}
\caption{\textbf{Wiring diagram of the measurement setup.}
}
\label{fig:figSI_WiringDiagram}
\end{figure}

After fabrication, the sample is placed in a copper microwave box and mounted to the mixing chamber stage of a Bluefors LD250 dilution refrigerator. A schematic of the measurement setup can be seen in Fig.~\ref{fig:figSI_WiringDiagram}. The sample is shielded from external magnetic fields by enclosing it in an inner aluminum shield and an outer shield made from high-permeability metal (Cryoperm). An external magnetic flux is applied through the SQUID of the transmon by a superconducting coil mounted below the sample within the magnetic shielding. The current is driven by a Stanford Research Systems CS580 voltage-controlled current source. Both qubit and resonator drives are generated by a Quantum Machines OPX+. The qubit drive, which needs to be below 2 GHz, is upconverted by a Rohde\&Schwarz SGS100A. Up and downconversion of the resonator drive are performed using a Quantum Machines Octave (frequency range of 2-18 GHz). Both qubit and resonator signals are attenuated and filtered in the dilution refrigerator. After reflection from the resonator input port, the resonator signal is first amplified by a SNAIL parametric amplifier (SPA) \cite{Frattini2018} fabricated in house. The SPA is enclosed in its own Cryoperm/aluminum magnetic shielding and biased by a coil, which is also controlled by a Stanford Research Systems CS580. The SPA is pumped at twice the resonator frequency using a Keysight PSG E8257D. Due to the very large resonator drive powers used in the experiments, the gain in most experiments is set to approximately 15 dB to limit the effects of amplifier saturation. Next, the resonator output passes through a Low Noise Factory (LNF) dual junction isolator, before it gets further amplified by an LNF high-electron mobility transistor (HEMT) amplifier at 4K. At room temperature, the signal is amplified a final time before demodulation and digitization in the Octave and OPX+, respectively. 
\\
\\
\\
\\
\\
\\

\section{Averaging of the Josephson potential in transmon circuits}
\label{sec:averaging}
In this section we derive analytically the conditions required to perform the averaging that lead to Hamiltonian~(1) of the main text. Then we extend this treatment to more generic transmon circuits, and show the ubiquity of the collapse and the inversion of the Josephson potential. 

\subsection{Flux-driven transmon with symmetric SQUID}
\begin{figure}[h]
\includegraphics[]{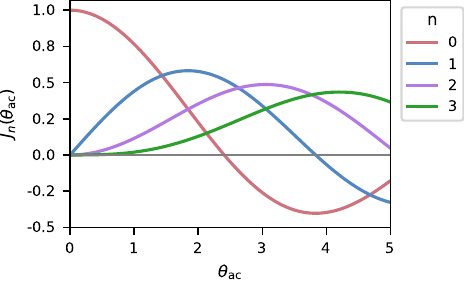}
\caption{\textbf{Bessel functions of the first kind.} }  
\label{fig:bessel}
\end{figure}

First we present in detail the system introduced in the concept part of the main text, that is a flux-driven transmon embedding a symmetric SQUID. Its Hamiltonian reads:
\begin{equation}
    \hat{\mathcal{H}} = 4E_C\big(\hat{N} - N_g\big)^2 - E_J \cos \frac{\pi\phi_\mathrm{ext}}{\phi_0} \cos\hat{\varphi}\;.
    \label{eq:H_fluxdriven_sym}
\end{equation}
In the presence of a monochromatic flux drive $\phiext(t) = \phidc + \phiac \sin\omega_dt$, we make use of the Jacobi-Anger expansion and decompose the Hamiltonian into $\hat{\mathcal{H}} = \hat{\mathcal{H}}_\mathrm{dc} + \hat{\mathcal{H}}_\mathrm{ac}$, where its steady part reads:
\begin{equation}
    \hat{\mathcal{H}}_\mathrm{dc} = 4E_C\big(\hat{N} - N_g\big)^2 - E_J J_0(\thetaac) \cos \thetadc \cos\hat{\varphi}\;,
    \label{eq:H_fluxdriven_sym_rwa}
\end{equation}
and the time-varying one:
\begin{equation}
    \hat{\mathcal{H}}_\mathrm{ac} = -2E_J\Bigg[ \cos \thetadc \sum_{n=1}^\infty J_{2n}(\thetaac)\cos (2n\omega_dt) + \sin \thetadc \sum_{n=1}^\infty J_{2n-1}(\thetaac)\cos \big((2n-1)\omega_dt)\big) \Bigg] \cos \hat{\varphi}\;,
\end{equation}
where $\theta_\mathrm{dc,ac} = \pi\phi_\mathrm{dc,ac}/\phi_0$ and $J_n$ are the n$^{th}$ order Bessel functions of the first kind (Fig.~\ref{fig:bessel}). Crucially the steady part corresponds to Hamiltonian~(1) of the main text, for which $\cos\thetadc=1$ at the upper sweet spot. 

Next we diagonalize the steady part into $\hat{\mathcal{H}}_\mathrm{dc} = \sum_k E_k \ket{k} \bra{k}$ where the eigenenergies $E_k$ and eigenstates $\ket{k}$ implicitly depend on $(N_g, \thetadc, \thetaac)$. Under the rotating-wave approximation (RWA) \cite{Mirrahimi2015}, the dynamics of the system at long times is accurately captured by the steady part of the Hamiltonian only, provided:
\begin{alignat}{2}
    \forall (n>0,k,k'), \quad &\big|E_J J_{2n}(\thetaac) \cos\thetadc \bra{k'}\cos\hat{\varphi}\ket{k}\big| &&\ll \big|E_k - E_{k'} \pm 2n\omega_d \big| \label{eq:cos_even}\\
    &\big|E_J J_{2n-1}(\thetaac) \sin\thetadc \bra{k'}\cos\hat{\varphi}\ket{k}\big| &&\ll \big|E_k - E_{k'} \pm (2n-1)\omega_d \big|\;. \label{eq:cos_odd}
\end{alignat}
While the right-hand sides define the multi-excitation resonances intrinsic to the transmon spectrum, these inequalities give the thresholds for their activation. The initial growth of $J_n(\thetaac)$ for $n>0$ demonstrates their susceptibility to strong drives (Fig.~\ref{fig:bessel}). When these conditions are fulfilled, the averaging theorem applies and one can recast the effects of the drive into a renormalization of the Josephson energy only. The collapse and inversion of the potential are captured by the oscillations of $J_0$ around zero. We believe that the observation of the simple spectroscopic features presented in the main text was allowed by the absence of activated multi-excitation resonances in our system. While the main text focused on the upper sweet spot $\cos\thetadc=1$, these features actually survive any dc-bias, as the existence of the collapse points are solely dictated by the zeros of $J_0(\thetaac)$.

\begin{figure}[h]
\includegraphics[width=\textwidth]{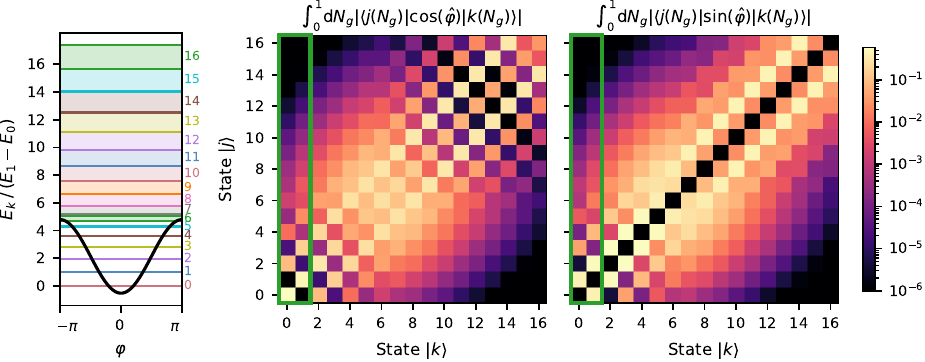}
\caption{\textbf{Transmon-parity protection.} The eigenenergies $E_k(N_g)$ and eigenstates $\ket{k(N_g)}$ of Hamiltonian~(\ref{eq:H_fluxdriven_sym_rwa}) are computed with $E_C/h=103$~MHz, $E_J/h=5.2$~GHz, and $\thetaac=\thetadc=0$. (Left) Transmon charge bands (color, state index $k$ on the right). The states $k<6$ are bounded by the Josephson potential well (black line). (Center) Charge-averaged matrix elements for the $\cos\hat{\varphi}$ operator. (Right) Same for the $\sin\hat{\varphi}$ operator. Focusing on the coupling of the computational states $k\in\{0,1\}$ to other states $j$ (green boxes), three regimes are identified. First, the coupling to states deep in the Josephson potential well $j<4$ obeys parity selection rules: $\cos\hat{\varphi}$ ($\sin\hat{\varphi}$) couples dominantly states with same (opposite) parity. Second, these rules are relaxed for states in the intermediate layer $4\leq j <11$. Third, the coupling to states way above the Josephson potential $j\geq11$ decays exponentially with $j$. Overall, the intermediate layer is the most detrimental to the system stability as it shows non-negligible couplings to the computational manifold and strong charge sensitivity.} 
\label{fig:matelts}
\end{figure}

These resonance conditions show clear selection rules. While the first set~(\ref{eq:cos_even}) captures transitions mediated by an even number of drive photons, the second one~(\ref{eq:cos_odd}) involves odd numbers. As such, tuning the static flux $\thetadc$ can select the \emph{photon parity} of the allowed processes. For both sets, the transitions within the transmon energy ladder are mediated by the $\cos\hat{\varphi}$ operator. Deep in the Josephson well it connects predominantly states with the same parity, but this \emph{transmon-parity} protection is rather weak as it does not hold for states near and above the top of the well (Fig.~\ref{fig:matelts}). Overall, at the upper sweet spot, a flux-driven transmon with a symmetric SQUID is protected against a wide range of multi-excitation resonances \cite{Chapple2025,Mori2025a,Mori2025b,Hazra2026}. It is thus the preferred operating point to image the collapse and inversion of the Josephson potential. Deviations from that sweet spot do not preclude its observation altogether, but the resulting activation of multi-excitation resonances may obscure the clean spectroscopic features presented in the main text.

\subsection{Flux-driven transmon with asymmetric SQUID}
Next we continue with the case of a transmon embedding an asymmetric SQUID \cite{Koch2007}:
\begin{equation}
    \hat{\mathcal{H}} = 4E_C\big(\hat{N} - N_g\big)^2 - E_J \cos\frac{\pi\phi_\mathrm{ext}}{\phi_0}\cos\hat{\varphi} - dE_J \sin\frac{\pi\phi_\mathrm{ext}}{\phi_0}\sin\hat{\varphi}\;,
    \label{eq:H_fluxdriven_asym}
\end{equation}
where now $E_J$ is the sum energy for the two junctions $E_J = E_{J1} + E_{J2}$ and $d=|E_{J1} - E_{J2}|/E_J$ is the asymmetry parameter. Similarly, under a time-dependent flux drive we write $\hat{\mathcal{H}} = \hat{\mathcal{H}}_\mathrm{dc} + \hat{\mathcal{H}}_\mathrm{ac}$ with:
\begin{equation}
    \hat{\mathcal{H}}_\mathrm{dc} = 4E_C\big(\hat{N} - N_g\big)^2 - E_J J_0(\thetaac) \cos \thetadc \sqrt{1 + d^2\tan^2 \thetadc}\cos\big(\hat{\varphi} - \varphi_0\big)\;,
\end{equation}
where $\tan\varphi_0 = d\tan\thetadc$, and:
\begin{align}
    \hat{\mathcal{H}}_\mathrm{ac} = 
    -2E_J&\Bigg[ \cos \thetadc \sum_{n=1}^\infty J_{2n}(\thetaac)\cos (2n\omega_dt) + \sin \thetadc \sum_{n=1}^\infty J_{2n-1}(\thetaac)\cos \big((2n-1)\omega_dt)\big) \Bigg] \cos \hat{\varphi} \nonumber \\
    -2dE_J&\Bigg[ \sin \thetadc \sum_{n=1}^\infty J_{2n}(\thetaac)\cos (2n\omega_dt) - \cos \thetadc \sum_{n=1}^\infty J_{2n-1}(\thetaac)\cos \big((2n-1)\omega_dt)\big) \Bigg] \sin \hat{\varphi} \;.
\end{align}
Interestingly, the renormalization of the Josephson energy imparted by the ac component of the flux drive fully decouples from its dc counterpart. In the presence of a finite asymmetry, while static flux tuning can only trim $E_J$ down to $dE_J$, a high-frequency drive can average it by a factor $J_0(\thetaac)$ which will go to zero regardless of the asymmetry. This is further evidence of the fundamental difference between tuning the static flux of a SQUID to $\phi_0/2$ and the dynamical renormalization of the potential.

Next, for the static part to faithfully depict the dynamical properties of the system, the previously defined conditions (\ref{eq:cos_even}) and (\ref{eq:cos_odd}) are supplemented by two new sets:
\begin{alignat}{2}
    \forall (n>0,k,k'), \quad &\big|dE_J J_{2n}(\thetaac) \sin\thetadc \bra{k'}\sin\hat{\varphi}\ket{k}\big| &&\ll \big|E_k - E_{k'} \pm 2n\omega_d \big| \label{eq:sin_even} \\
    &\big|dE_J J_{2n-1}(\thetaac) \cos\thetadc \bra{k'}\sin\hat{\varphi}\ket{k}\big| &&\ll \big|E_k - E_{k'} \pm (2n-1)\omega_d \big|\;. \label{eq:sin_odd}
\end{alignat}
First, the presence of a finite asymmetry breaks the photon-parity protection that was granted to the sweet spots. Second, through the $\sin\hat{\varphi}$ matrix elements it breaks the already fragile transmon-parity selection rule (Fig.~\ref{fig:matelts}). Yet, just like dc bias away from the upper sweet
spot, it does not change the definition of the collapse points, but simply increases the number of spurious processes that may hinder its observation. 

\subsection{Charge-driven transmon}
\label{sec:charge_driven_trm}
Finally we consider a fixed-frequency transmon charge-driven through a capacitively coupled port:
\begin{equation}
    \hat{\mathcal{H}} = 4E_C\big(\hat{N} - N_g\big)^2 - E_J\cos\hat{\varphi} - \hbar\varepsilon_d \hat{N} \cos \omega_d t\;,
\end{equation}
where $\varepsilon_d$ is the strength of the drive with frequency $\omega_d$. Following refs. \cite{Lescanne2019,Verney2019}, we transform this charge drive into a phase drive in the moving frame defined by the unitary $U_\theta=e^{i\theta_0\sin\omega_d t\hat{N}}$ where $\theta_0=\varepsilon_d/\omega_d$. The displaced Hamiltonian $\hat{\mathcal{H}}_\theta = U_\theta \hat{\mathcal{H}} U_\theta^\dagger - i\hbar U_\theta \dot{U}_\theta^\dagger$ reads:
\begin{equation}
    \hat{\mathcal{H}}_\theta = 4E_C\big(\hat{N} - N_g\big)^2 - E_J\cos \big(\hat{\varphi} + \theta_0\sin\omega_dt\big)\;.
\end{equation}
We can decompose it into $\hat{\mathcal{H}}_\theta = \hat{\mathcal{H}}_{\theta,\mathrm{dc}} + \hat{\mathcal{H}}_{\theta,\mathrm{ac}}$ where:
\begin{equation}
    \hat{\mathcal{H}}_{\theta,\mathrm{dc}} = 4E_C\big(\hat{N} - N_g\big)^2 - E_J J_0(\thetaac) \cos\hat{\varphi}\;,
\end{equation}
and:
\begin{equation}
    \hat{\mathcal{H}}_{\theta,\mathrm{ac}} = -2E_J \Bigg[ \sum_{n=1}^\infty J_{2n}(\thetaac)\cos (2n\omega_dt) \cos\hat{\varphi} + \sum_{n=1}^\infty J_{2n-1}(\thetaac)\cos \big((2n-1)\omega_dt)\big) \sin\hat{\varphi} \Bigg]\;.
\end{equation}
Just like before, one can safely average out the time-dependent contribution if:
\begin{alignat}{2}
    \forall (n>0,k,k'), \quad &\big|E_J J_{2n}(\thetaac) \bra{k'}\cos\hat{\varphi}\ket{k}\big| &&\ll \big|E_k - E_{k'} \pm 2n\omega_d \big| \\
    &\big|E_J J_{2n-1}(\thetaac) \bra{k'}\sin\hat{\varphi}\ket{k}\big| &&\ll \big|E_k - E_{k'} \pm (2n-1)\omega_d \big|\;.
\end{alignat}
In this case, no more selection rules hold, and the charge-driven transmon is maximally exposed to multi-excitation resonances. Yet, the renormalization of the Josephson energy still has the same form, demonstrating the ubiquity of the collapse and inversion in transmon circuits.

\subsection{Beyond transmons}
The collapse and inversion of the Josephson potential is less a feature specific to transmon circuits than a generic regime of Josephson junctions under strong and high-frequency drives. Ultimately, it is the product of the junction cosinusoidal energy-phase relation under a periodic excitation. One can imagine replicating this effect in a fluxonium-type circuit. There, the collapse of the potential would coincide with the emergence of a dynamically linearized oscillator, akin to the Frozonium regime \cite{Lewellen2026}. Alternatively, one may wonder what happens in the presence of high-order Josephson harmonics $E_{Jn}\cos n\hat{\varphi}$. There, every harmonic will collapse at a different drive amplitude set by $J_0(n\thetaac)=0$. This asynchronous renormalization of the Josephson harmonics is only expected to happen in junctions natively featuring higher harmonics. On the other hand, the $\cos 2\hat{\varphi}$ contribution emerging from the inductive loading of tunnel junctions will collapse together with the bare junction potential \cite{Kim2026}.

\section{Circuit Design}
The observation of the collapse and inversion of the Josephson potential requires the circuit to be resilient to multi-excitation resonances. This condition is more easily met for a flux-driven transmon embedding a symmetric SQUID (section~\ref{sec:averaging}). In the first part, we present a class of circuits that can practically realize this paradigmatic driving scheme. In the second part, we model the device studied in the main text and perform circuit quantization. In the third part, we detail the design of the resonator. In the fourth part, we present the Floquet simulations of the system. And finally in the fifth part, we break down the readout mechanisms coming out of the cosine-cosine coupling.

\subsection{Cosine-Cosine Coupling in Transmon Circuits}
In section~\ref{sec:averaging}, we highlighted the transmon driving scheme that is the least susceptible to multi-excitation resonances: the flux drive at the sweet spot. It couples the $\cos\hat{\varphi}$ operator to $\cos\thetaac$, where $\thetaac$ is the equivalent phase drive triggered by the excitation field (Eq.~(\ref{eq:H_fluxdriven_sym})). If one replaces the stiff drive by the quantized field living in a resonator $\thetaac\rightarrow \hat{\theta}$, this realizes the \emph{cosine-cosine} coupling that recently drew a lot of attention in the context of qubit readout \cite{Dassonneville2020, Salunkhe2025, Mori2025a, Mori2025b, Chapple2025, Hazra2025, Hazra2026}. Indeed, while the linear dipolar interaction at the heart of most cQED designs mediates a cross-Kerr interaction only in the perturbative sense, this nonlinear cosine-cosine coupling is the primitive of a non-perturbative cross-Kerr interaction (section~\ref{sec:readout}). It allows the implementation of standard dispersive readout, with an added resilience to strong readout powers. Over the past few years, a rather diverse class of circuits has implemented this nonlinear coupling between a transmon and a resonator. In this part, we briefly compare their designs, which will help us motivate the device presented in the main text and detailed in the next part (section~\ref{sec:circuit_quant}).
\\

\begin{figure}[b]
    \begin{tabular}{c c c c}
        \textbf{Photon-Pressure} \cite{Potts2025} & \textbf{Transmon Molecule} \cite{Dassonneville2020,Mori2025a,Mori2025b} & \textbf{Dimon} \cite{Hazra2025,Hazra2026} & \textbf{Quantromon} \cite{Salunkhe2025} \\
        G. Steele (Delft) & O. Buisson (Grenoble) & M. Devoret (Yale) & R. Vijay (Mumbai) \\ \\ 
        \includegraphics{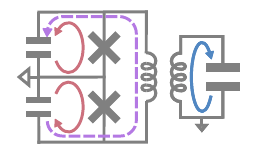}
        &\includegraphics{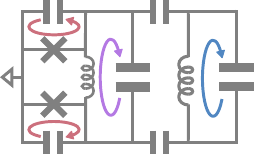}\hspace{0.1cm}
        &\hspace{0.1cm}\includegraphics{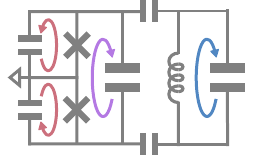}
        &\includegraphics{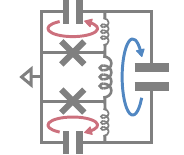}
        \vspace{0.2cm} \\
        \textbf{2+1 Modes} & \textbf{3 Modes} & \textbf{3 Modes} & \textbf{2 Modes} \\
        \textcolor{red}{Transmon (nonlinear)}
        &\textcolor{red}{Transmon (nonlinear)}
        &\textcolor{red}{Transmon (nonlinear)}
        &\textcolor{red}{Transmon (nonlinear)} \\
        \textcolor{blu}{Readout (linear)} 
        &\textcolor{blu}{Readout (linear)} 
        &\textcolor{blu}{Readout (linear)} 
        &\textcolor{blu}{Readout (weakly nonlinear)}  \\
        \textcolor{pur}{Spurious (weakly nonlinear)}
        &\textcolor{pur}{Mediator (weakly nonlinear)}
        &\textcolor{pur}{Mediator (nonlinear)}
        &
        \vspace{0.2cm} \\
        \textbf{Flux Loop} & \textbf{Flux Loop} & \textbf{No Flux Loop} & \textbf{Flux Loop}
        \vspace{0.2cm} \\
        \textbf{Couplings} & \textbf{Couplings} & \textbf{Couplings} & \textbf{Couplings} \\
        \textcolor{red}{T}-\textcolor{blu}{R}: cosine-cosine (flux)
        &\textcolor{red}{T}-\textcolor{pur}{M}: cosine-cosine (phase)
        &\textcolor{red}{T}-\textcolor{pur}{M}: cosine-cosine (phase)
        &\textcolor{red}{T}-\textcolor{blu}{R}: cosine-cosine (phase)\\
        \textcolor{pur}{S}-\textcolor{blu}{R}: dipolar (flux)
        &\textcolor{pur}{M}-\textcolor{blu}{R}: dipolar (charge)
        &\textcolor{pur}{M}-\textcolor{blu}{R}: dipolar (charge)
        & \\
        \textcolor{red}{T}-\textcolor{pur}{S}: cosine-cosine (phase)
        &
        &
        &
    \end{tabular}
    \caption{\textbf{Zoology of the circuits implementing a cosine-cosine coupling between a transmon and a readout mode.} Colored arrows represent the current flowing through the bare circuit modes. The degree of linearity of each mode is categorized between: nonlinear (self-Kerr $\gg 1$~MHz), weakly nonlinear (self-Kerr $\sim 1$~MHz) or linear (self-Kerr $\ll1$~MHz). Here, we consider circuits symmetric about the horizontal axis. Any asymmetry in the junction or capacitive energies would lead to extra dipolar couplings between the transmon and the other modes.}
    \label{fig:zoology}
\end{figure}

\noindent\textbf{Photon-Pressure}\hspace{0.2cm} We begin in a non-chronological order with the Photon-Pressure architecture, imagined by the Delft team to replicate the optomechanical coupling in circuit-QED. It features a symmetric SQUID loaded by an inductor. A folded capacitance matrix shapes the transmon mode as the superposition of two counter-propagating currents in the junctions. The readout mode is hosted on the same chip in a mutually coupled resonator. This coupling scheme closely follows the original flux-drive intuition, as the current running through the resonator generates flux in the SQUID loop. The transmon and the resonator are cosine-cosine coupled through flux, and the interaction strength is set by the participation of the mutual in the total SQUID inductance. This circuit features a third spurious mode, also cosine-cosine coupled to the transmon, and inductively coupled to the readout mode. \\

\noindent\textbf{Transmon Molecule}\hspace{0.2cm} We continue with the Transmon Molecule, a circuit designed by the Grenoble team and that initiated the search for non-perturbative cross-Kerr interactions in circuit-QED. It features a symmetric SQUID loaded by a coupling (super)inductor. The symmetric mode of the SQUID, loaded by the capacitors parallel to the junctions, is the transmon mode. The differential mode of the SQUID, loaded by the capacitor in parallel with the inductor, is a mediator mode, cosine-cosine coupled to the transmon. The interaction strength is set by the participation ratio of the coupling inductor in the total SQUID inductance. If we consider the magnetic field stored in the inductor to ``self-bias" the SQUID loop, we recover the flux-drive intuition. Yet this architecture goes beyond, as a purely kinetic inductor can also mediate the coupling through a phase bias of the SQUID loop. The mediator is capacitively coupled to the readout mode which is hosted in an auxiliary 3D cavity. Near resonance they hybridize, so that the transmon ends up cosine-cosine coupled through phase and charge to the resulting polariton modes. \\

\noindent\textbf{Dimon}\hspace{0.2cm} The Dimon circuit was designed by the Yale team, and can be thought of as a Transmon Molecule free of the superconducting loop. It relies on the same folded geometry as the Photon-Pressure device, with a similar transmon mode. As for the mediator, it consists of the superposition of two co-propagating currents in the junctions. As a consequence, it is cosine-cosine coupled to the transmon through phase, and its participation is maximized. The price to pay is that the mediator mode is now as nonlinear as the transmon, and the polaritons emerging from the coupling with the readout mode also inherit a sizeable nonlinearity. Interestingly this architecture completely challenges the original flux-drive intuition, as it does not even feature a superconducting loop. \\

\noindent\textbf{Quantromon}\hspace{0.2cm} The Quantromon circuit was designed by the Mumbai team, and up to minor modifications it corresponds to a Transmon Molecule chip inserted in a 3D waveguide. As a consequence, there is no auxiliary 3D mode, and what was previously defined as the mediator plays the role of the readout mode. The waveguide architecture has a great potential in preventing spurious modes from interacting with the circuit if the latter is placed well below cutoff, which was not the case in this experiment. This circuit was also discussed in theory by the team of A. Blais \cite{Chapple2025}. \\

All these circuits implement a cosine-cosine coupling between transmon and readout, thus limiting the amount of intrinsic multi-excitation resonances. However, since every additionally coupled mode can mediate extra spurious processes \cite{Hazra2026}, these circuits are expected to perform differently under strong drives. Specifically, the presence of a third mode in the Photon-Pressure, Transmon Molecule and Dimon circuits must limit their dynamic range. One can also add the enhanced nonlinearity of the mediator in the Dimon, or the all-to-all coupling matrix of the Photon-Pressure as extra limiting factors. Overall, the two-mode design of the Quantromon seems like the best strategy for the observation of the collapse and inversion of the Josephson potential.

\subsection{Circuit Modeling}
\label{sec:circuit_quant}
The device presented in the main text can be thought of as a Quantromon integrated in a 2D architecture. While 2D circuit-QED is more suscpetible to spurious modes than its 3D-waveguide counterpart, our device leverages grounding points to prevent their appearance. In this part, we will first describe the challenge in giving a circuit description to our fully grounded device, and then perform circuit quantization of an effective model. Finally, we will connect the effective parameters to the actual device.

\subsubsection{Equivalent and Effective Circuits}
\begin{figure}[t]
\includegraphics[]{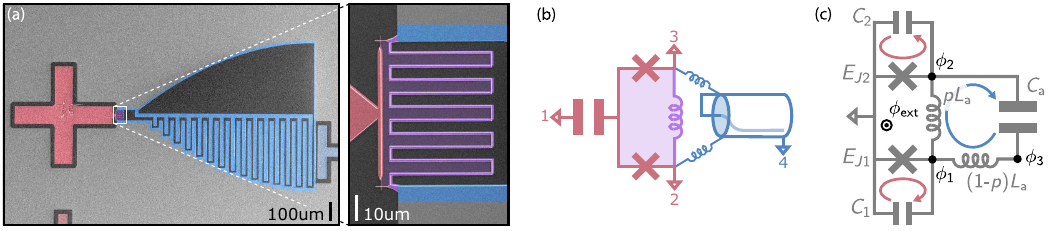}
\caption{\textbf{Circuit analysis.} (a) False-colored scanning electron micrograph of the device, with the transmon island and junctions in salmon, the resonator in blue and the coupling inductor in purple. (b) Equivalent circuit of the device, where multiple grounds are defined as a reminder of its distributed nature. The asymmetric transmission line represents the one-sided interdigitated capacitor of the quarter-wavelength resonator. (c) Effective lumped-element circuit of the transmon and resonator modes capturing the low-energy properties of the device (see detail in text). This circuit has the same mode structure as the Quantromon (Fig.~\ref{fig:zoology}).}  
\label{fig:circuitquant}
\end{figure}

In Fig.~\ref{fig:circuitquant}(a), we show the device picture, which features a flux-tunable transmon in a grounded geometry \cite{Barends2013}, and a quarter-wavelength resonator shunted to ground in a hook below the transmon. The cosine-cosine coupling is powered by the phase bias imparted by the fraction $p_I$ of the resonator current that flows through the side arm of the SQUID (purple meander). While the grounded nature of the resonator naturally sacrifices some of the coupling current ($p_I<1$), it also pushes up in frequency the higher harmonics that could mediate multi-excitation resonances. As for its unconventional interdigitated-capacitor design, it is motivated in section~\ref{sec:resonator_design}. In the end, the translation of the device into an equivalent circuit is made quite complex by the combination of the hook asymmetry, and the galvanic connection of the coupling inductor to the ground plane at both ends. In Fig.~\ref{fig:circuitquant}(b), we show our best attempt, which features multiple ground points with different potentials. An alternative description with a single ground would require connecting all those points with additional circuit elements that would ultimately influence the bare circuit modes. Being confronted with the very distributed nature of our design, we resort to finite-element simulation to elucidate its mode structure in a first step, and write an effective circuit model in a second step. 

We simulate our design using Ansys HFSS where the two junctions are encoded as lumped elements with identical inductance $L_J=55$~nH. The low-frequency circuit modes are found to be the lumped mode of the transmon island around $\omega_{01}/2\pi=1.86$~GHz, and the fundamental of the quarter-wavelength resonator around $\omega_\mathrm{a}/2\pi=8.07$~GHz. The next one is the first excited mode of the quarter-wavelength resonator, with a frequency greater than 20~GHz, way above the lowest package mode (around 16~GHz). This motivates the description of the low-energy properties of our device using the effective lumped-element circuit shown in Fig.~\ref{fig:circuitquant}(c).

\subsubsection{Circuit Quantization}
The effective circuit features two Josephson junctions with energies $E_{J1,2}$ arranged in a SQUID threaded by a magnetic flux $\phi_\mathrm{ext}$. The corresponding junction inductances are $L_{J1,2} = (\phi_0/2\pi)^2/E_{J1,2}$ where $\phi_0$ is the flux quantum. The junctions are loaded by parallel capacitors $C_{1,2}$, and the side arm of the SQUID features an inductor $p\La$ with $0<p<1$. It is loaded in parallel with an inductor $(1-p)\La$ in series with a capacitor $\Ca$. At this stage, we emphasize that $p\La$ is not the inductance of the meander of the actual device, but rather the coupling inductance of the effective model. The ratio $p$ is the dilution factor of the phase cosine-cosine coupling. We describe this circuit in terms of the generalized node-flux coordinates $\phi_{1,2,3}$. Its classical Lagrangian reads $\mathcal{L} = \mathcal{T} - \mathcal{U}$ where:
\begin{align}
    \mathcal{T} &= \frac{1}{2}\Ca\big(\dot{\phi}_2-\dot{\phi}_3\big)^2 + \frac{1}{2}C_1\dot{\phi}_1^2 + \frac{1}{2}C_2\dot{\phi}_2^2 \;, \\
    \mathcal{U} &= \frac{1}{2p\La} \big(\phi_1 - \phi_2 + \phiext\big)^2 + \frac{1}{2(1-p)\La}\big(\phi_1 - \phi_3\big)^2 - E_{J1}\cos\varphi_1 - E_{J2}\cos\varphi_2\;,
\end{align}
and we used the reduced flux variables $\varphi_k = (2\pi/\phi_0)\phi_k$. Next we define the new variables $\phi_\Sigma = \frac{1}{2}(\phi_1 + \phi_2)$, $\phi_\Delta = \frac{1}{2}(\phi_1 - \phi_2) + \tfrac{1}{2}\phiext$, and $\phia = \phi_2 - \phi_3 - \frac{1}{2}\phiext$ so that we find up to irrelevant constant terms:
\begin{align}
    \mathcal{T} &= \frac{1}{2}\Ca\big(\dot{\phi}_\mathrm{a} + \tfrac{1}{2} \dot{\phi}_\mathrm{ext} \big)^2 + \frac{1}{2}\big(C_1 + C_2\big)\big(\dot{\phi}_\Sigma^2 + \dot{\phi}_\Delta^2 - \dot{\phi}_\mathrm{ext} \dot{\phi}_\Delta \big) + \frac{1}{2}\big(C_1 - C_2\big)\big(2\dot{\phi}_\Sigma \dot{\phi}_\Delta - \dot{\phi}_\mathrm{ext} \dot{\phi}_\Sigma \big)\;, \\
    \mathcal{U} &= \frac{1}{2p\La} \big(2\phi_\Delta\big)^2 + \frac{1}{2(1-p)\La}\big(2\phi_\Delta + \phia\big)^2 - E_{J1}\cos\big(\varphi_\Sigma + \varphi_\Delta - \tfrac{1}{2}\varphi_\mathrm{ext}\big) - E_{J2}\cos \big(\varphi_\Sigma - \varphi_\Delta + \tfrac{1}{2}\varphi_\mathrm{ext}\big) \;,
\end{align}
In the following, we limit the discussion to identical capacitances $C_1=C_2$, which cancels the charge dipolar-coupling between $\phi_\Sigma$ and $\phi_\Delta$. We also consider solely static offset fluxes, $\phi_{ext}=\phidc$, which cancels the effective charge drives on all modes. 

The symmetric SQUID mode $\phi_\Sigma$ captures the transmon, while $\phia$ is the readout mode. As for $\phi_\Delta$, it is the differential SQUID mode which, considering the grounding pattern of the actual circuit, needs to be very high-frequency. As a consequence, we employ a separation of timescales and write a slow Lagrangian for the modes $(\phi_\Sigma, \phia)$ with $\phi_\Delta$ ``frozen" \cite{Lu2023}. The equation of motion for the $\phi_\Delta$ mode reads:
\begin{equation}
    (C_1 + C_2) \ddot{\phi}_\Delta - \frac{4}{p\La}\phi_\Delta - \frac{2}{(1-p)\La}\big(2\phi_\Delta+\phia\big) = 0\;,
\end{equation}
where we neglect the junction contribution in the limit $L_{J1,2}\gg\La$. In the steady state, we find $\phi_\Delta = -\frac{1}{2}p\phia$, so that the slow Lagrangian reads $\mathcal{L}_\mathrm{slow} = \mathcal{T}_\mathrm{slow} - \mathcal{U}_\mathrm{slow}$ with:
\begin{align}
    \mathcal{T}_\mathrm{slow} &= \frac{1}{2} \big(\Ca + \tfrac{1}{4}p^2C_\Sigma\big) \dot{\phi}_\mathrm{a}^2 + \frac{1}{2} C_\Sigma \dot{\phi}_\Sigma^2 \;, \\
    \mathcal{U}_\mathrm{slow} &= \frac{1}{2\La} \phia^2 - E_{J\Sigma} \cos\big(\tfrac{1}{2}p\varphia + \tfrac{1}{2}\varphidc\big) \cos\varphi_\Sigma - dE_{J\Sigma} \sin\big(\tfrac{1}{2}p\varphia + \tfrac{1}{2}\varphidc\big) \sin\varphi_\Sigma \;,
\end{align}
where $C_\Sigma=C_1+C_2$, $E_{J\Sigma}=E_{J1}+E_{J2}$ and $d=(E_{J2}-E_{J1})/(E_{J2}+E_{J1})$.
Next, we introduce the canonical momenta $Q_\mathrm{a,\Sigma} = \partial \mathcal{L}_\mathrm{slow} / \partial \dot{\phi}_{\mathrm{a},\Sigma}$, write the classical Hamiltonian $\mathcal{H} = \sum_i \dot{\phi}_i Q_i - \mathcal{L}_\mathrm{slow}$, and promote all variables to quantum operators:
\begin{equation}
    \hat{\mathcal{H}} = \frac{\hat{Q}_\mathrm{a}^2 }{2(\Ca + \tfrac{1}{4}p^2C_\Sigma)}+ \frac{\hat{\phi}_\mathrm{a}^2 }{2\La}+ 4E_{C\Sigma}\big( \hat{N}_\Sigma - N_g\big)^2 - E_{J\Sigma} \cos\frac{\varphidc + p\hat{\varphi}_\mathrm{a}}{2} \cos\hat{\varphi}_\Sigma - d E_{J\Sigma} \sin\frac{\varphidc + p\hat{\varphi}_\mathrm{a}}{2} \sin\hat{\varphi}_\Sigma \;,
\end{equation}
where $E_{C\Sigma}=e^2/2C_\Sigma$, and $\smash{\hat{N}_\Sigma} = \hat{Q}_\Sigma/2e$ describes the excess number of Cooper pairs on the transmon island, with $[\hat{\varphi}_\Sigma, \hat{N}_\Sigma]=i$. The readout operators obey the commutation relation $[\hat{\phi}_\mathrm{a}, \hat{Q}_\mathrm{a}]=i\hbar$. In a final step, we introduce $\ha$ the annihilation operator for the resonator, following the commutation relation $[\ha,\had]=1$, so that:
\begin{equation}
    \hat{\mathcal{H}} = \hbar\omegaa \had\ha+ 4E_{C\Sigma}\big( \hat{N}_\Sigma - N_g\big)^2 - E_{J\Sigma} \cos\frac{\varphidc + p\varphi_{zpf}(\ha+\had)}{2} \cos\hat{\varphi}_\Sigma - d E_{J\Sigma} \sin\frac{\varphidc + p\varphi_{zpf}(\ha+\had)}{2} \sin\hat{\varphi}_\Sigma \;,
    \label{eq:full_hamiltonian}
\end{equation}
where:
\begin{equation}
    \omegaa = \frac{1}{\sqrt{\La(\Ca + \tfrac{1}{4}p^2C_\Sigma)}}\;,
    \qquad
    \varphi_{zpf} = \sqrt{\pi\frac{Z_\mathrm{a}}{R_K}}\;,
    \qquad
    Z_\mathrm{a} = \sqrt{\frac{\La}{\Ca+\frac{1}{4}p^2C_\Sigma}}\;,
\end{equation}
with $R_K=h/(2e)^2$ the resistance quantum. We recognize here the Hamiltonian previously derived in ref. \cite{Chapple2025}, up to the participation ratio $p$ that dilutes the cosine-cosine coupling. How does it relate to the actual device?

\subsubsection{Dilution Factor}
In our circuit, the coupling is powered by the phase bias imparted by the readout resonator across the SQUID inductor: $\varphi_\mathrm{bias} =  (2\pi/\phi_0)L_S I_S$. Here $L_S$ is the coupling inductance of the actual device, which amounts to a fraction $p_L=L_S/\La$ of the total mode inductance $\La$. Similarly $I_S$ is the fraction of the readout current flowing through that inductor, and it amounts to a fraction $p_I=I_S/I_{zpf}$ of the zero-point current fluctuations of the readout mode $I_{zpf}=(\phi_0/2\pi) \times \varphi_{zpf}/\La$. Overall, we find:
\begin{equation}
    \varphi_\mathrm{bias}  =  p \times \varphi_{zpf}
    \qquad
    \text{where}
    \qquad
    p = p_L \times p_I\;.
    \label{eq:participation_ratios}
\end{equation}
While for the effective model $p_I$ was implicitly set to 1, in the real device, a significant fraction of the resonator current does not contribute to the coupling.

\subsection{Resonator Design}
\label{sec:resonator_design}

\begin{figure}[h]
\includegraphics[width=\textwidth]{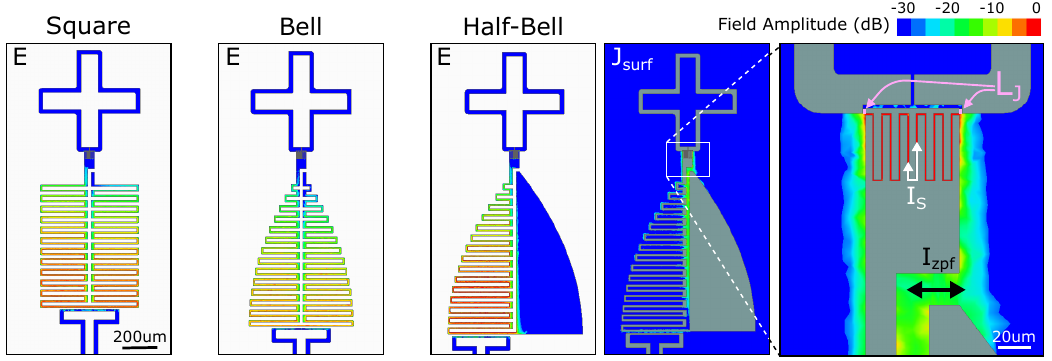}
\caption{\textbf{Resonator Design Evolution.} For the Square (left), Bell (center), and Half-Bell (right) geometries we plot the electric field distribution (E) in the gaps of the coplanar waveguides (metalized areas in white). For the Half-Bell geometry, we also plot the surface current-density field distribution ($\mathrm{J_{surf}}$) on the metalized areas (substrate in green), with a zoom-in on the coupler region. The transmon junctions are encoded as two lumped elements (pink). The colormap, normalized to the maximum field amplitude, is common to all plots.}  
\label{fig:design}
\end{figure}

In this part, we present the evolution of the resonator design, comparing three geometries (Fig.~\ref{fig:design}): the Square, the Bell and the Half-Bell. The optimization follows two guidelines: maximizing the cosine-cosine coupling, and minimizing the spurious dipolar coupling between the transmon and the resonator. We will first present both efforts, and then comment the results.

\subsubsection{Maximizing cosine-cosine coupling}
Eq.~(\ref{eq:participation_ratios}) deceivingly suggests that maximizing the phase fluctuations of the readout resonator will maximize the phase bias. In reality, the cosine-cosine coupling strength is set by the phase fluctuations of the resonator \emph{on the SQUID inductance}, and is better captured by the equation:
\begin{equation}
    \varphi_\mathrm{bias} = \frac{2\pi}{\phi_0} \times p_I \times L_S I_{zpf}\;.
\end{equation}
The route to maximizing the phase bias is now clear. First, we maximize the SQUID inductance by meandering a 450~$\mu$m-long and 1~$\mu$m-wide aluminum wire, for which we estimate $L_S=150$~pH using Ansys Q3D. Second, we maximize the zero-point current fluctuations of the readout resonator by minimzing its impedance. Indeed, for a quarter-wavelength resonator of frequency $\omegaa$ made out of a transmission line with characteristic impedance $Z_0$, one finds:
\begin{equation}
    I_{zpf} = \omegaa\sqrt{\frac{2\hbar}{\pi Z_0}}.
\end{equation}
As a consequence, we start the optimization with the Square geometry that features a fishbone transmission line with $Z_0=11$~$\Omega$ (estimated using Ansys HFSS). Third, we tweak the symmetry of the resonator to maximize the current participation ratio. In a symmetric geometry -- such as the Square or the Bell -- the current going out of the hook prefers the low impedance return-path that does not go through the coupling inductor, leading to $p_I<0.5$. In the Half-Bell, we break the symmetry of the transmission line to favor the coupling path and increase $p_I$. 

We quantify the magnitude of the zero-point current fluctuations in the resonator mode and in the SQUID inductance using Ansys HFSS. Specifically we reconstruct the currents at the locations marked in Fig.~\ref{fig:design} by integrating the magnetic field distribution in closed contours.

\subsubsection{Minimizing dipolar coupling}
As shown in section~\ref{sec:charge_driven_trm}, dipolar coupling is especially harmful regarding multi-excitation resonances. Charge-dipolar coupling appears when the resonator loads the transmon asymmetrically (section~\ref{sec:circuit_quant}). Inductive-dipolar coupling can also arise in the presence of a finite SQUID asymmetry (Eq.~(\ref{eq:full_hamiltonian})). We implement the Bell and Half-Bell geometries to minimize the charge-dipolar coupling by reducing the exposure of the transmon to the electric field stored in the resonator. 

We quantify the strength of the residual dipolar interaction $g_x$ by tracking avoided crossings using AnsysHFSS. Specifically, by sweeping the junction inductances until the transmon mode anticrosses with the readout mode, we can extract the frequency splitting $\Delta\omega$. Because in the actual device the transmon and resonator are far detuned, we relate this resonant splitting to the residual dipolar interaction using: $g_x=\frac{1}{2}\Delta\omega\times\sqrt{\omega_{01}/\omega_\mathrm{a}}$ \cite{Blais2021}.

\subsubsection{Results}
In Table~\ref{fig:design_result}, we present the results of the simulation. While the current participation ratio approaches 0.5 in the Half-Bell geometry, demonstrating a 40\% increase as compared to the Square, the current running through the SQUID only increases by 14\%. This moderate improvement can be understood by the joint increase of the resonator impedance when removing one side of the fishbone, which decreases the zero-point current fluctuations of the resonator by 20\%. Finally, we find that reducing the capacitive exposure of the resonator to the transmon in going from the Square to the Bell geometry reduces the dipolar coupling by a factor greater than 2. This reduction is almost preserved with the Half-Bell. In the end, the main improvement brought by this unconventional geometry -- besides its unique look -- lies in the reduction by a factor of 2 of the residual dipolar coupling, highly beneficial to limit the impact of multi-excitation resonances (section~\ref{sec:charge_driven_trm}).

\begin{table}[h]
\caption{\textbf{Simulated parameters for all three geometries.}}
\begin{tabular}{r c c c}
    \toprule
        & \;\; \textbf{Square} \;\; & \;\; \textbf{Bell} \;\; & \;\; \textbf{Half-Bell} \;\; \\
      $p_I=I_S/I_{zpf}$ & 0.35 & 0.33 & 0.49 \\
      $I_{zpf}$ [nA] & 123 & 120 & 99 \\
      $I_S$ [nA] & 43 & 40 & 49 \\
      $g_x/2\pi$ [MHz] & 7.9 & 3.4 & 4.1 \\
    \bottomrule
\end{tabular}  
\label{fig:design_result}
\end{table}

\subsection{Floquet Analysis}
\label{sec:SI_FloquetAnalysis}

In this part we present the Floquet analysis of the device studied in the main text and described in section~\ref{sec:circuit_quant}. We begin with a presentation of the semiclassical approximation underlying the Floquet treatment. Then we quantitatively analyze the degree of activation of the multi-excitation resonances encountered in the Floquet spectrum. Finally we compare the results from the Floquet theory with the averaged model.

\subsubsection{Semiclassical approximation}
In the presence of a strong resonant drive, the resonator field is displaced into a large coherent state $\alpha(t)=\sqrt{n}\sin\omega_dt$ with $n$ its mean occupation and $\omega_d$ the drive frequency. In the correspondence limit, we can neglect the resonator quantum fluctuations and write a semiclassical approximation of Hamiltonian~(\ref{eq:full_hamiltonian}):
\begin{equation}
    \hat{\mathcal{H}}_\mathrm{sc}(t) = 4E_{C}\big( \hat{N} - N_g\big)^2 - E_J \cos\big(\tfrac{1}{2}\varphidc + \thetaac\sin\omega_dt \big)  \cos\hat{\varphi} - d E_J \sin\big(\tfrac{1}{2}\varphidc + \thetaac\sin\omega_dt \big) \sin\hat{\varphi} \;,
    \label{eq:sc_hamiltonian}
\end{equation}
where $\thetaac=p\varphi_{zpf}\sqrt{n}$ is the phase-drive amplitude, and we dropped the $\Sigma$ index of the transmon mode for the sake of simplicity. We recover Hamiltonian~(\ref{eq:H_fluxdriven_asym}), but while in section~\ref{sec:averaging} we resorted to the rotating-wave approximation to elucidate multi-excitation resonances, here we will use the exact Floquet method \cite{Grifoni1998, Dumas2024}.

Taking advantage of the periodicity of the Hamiltonian $\hat{\mathcal{H}}_\mathrm{sc}(t) = \hat{\mathcal{H}}_\mathrm{sc}(t+T)$ with $T=2\pi/\omega_d$, we can expand the solutions of the Schrodinger equation on the basis of Floquet states $\ket{\Psi_k(n,N_g;t)}=e^{-\text{i}\varepsilon_k(n,N_g)t}\ket{k(n,N_g;t)}$. The Floquet modes $\ket{k(n,N_g;t)}$, or in other words the dressed transmon states, are $T$-periodic. This time dependence captures the fast dynamics imparted by the drive. In turn, the corresponding Floquet quasi-energies $\varepsilon_k(n,N_g)$ are defined modulo $\omega_d$. They constitute the Floquet spectrum that captures the slow properties of the system. The replicas of each level by $m\times\omega_d$ with $m\in\mathbb{Z}$ translate the presence of the drive field, which is an infinite reservoir of drive photons that can exchange energy quanta with the transmon. As a consequence, a single Brillouin zone $[0,\omega_d]$ is sufficient to characterize the Floquet spectrum, where resonances between two levels $i$ and $j$ are actually multi-excitation resonances satisfying $\varepsilon_i(n,N_g)-\varepsilon_j(n,N_g)=m\omega_d$. In the following we will analyze the resilience of our system to strong resonator drives by computing its Floquet spectrum using QuTiP \cite{qutip5}.

\subsubsection{Resilience to multi-excitation resonances}

\begin{figure}[h]
\includegraphics[]{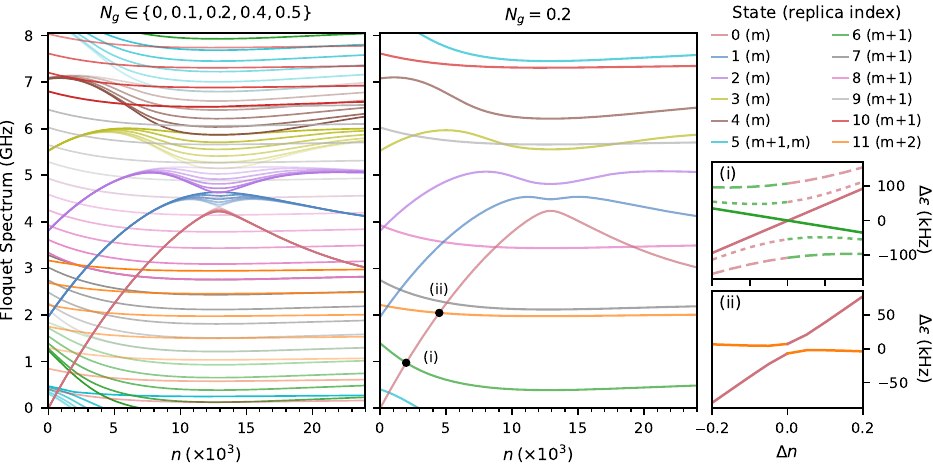}
\caption{\textbf{Resilience to multi-excitation resonances}. Floquet simulation of Hamiltonian~(\ref{eq:sc_hamiltonian}) computed with $E_C/h=103$~MHz, $E_J/h=5.22$~GHz, $\omega_d/2\pi=8.054$~GHz, $d=0$, $\varphidc=0$ and $p\varphi_{zpf}=0.0216$. (Left) Floquet spectrum versus drive strength quantified by the occupation of the resonator (x-axis), for multiple values of the offset charge from 0 to 0.5 (dark to light colors). (Center) Focus on $N_g=0.2$ for which the replica indices of the energy levels are specified in the legend. Note that level 5 changes index around $n\approx3\times10^3$. (Insets) Close-ups on two multi-excitation resonances for $d=0$ (solid lines), $d=0.01$ (short dashed lines), and $d=0.02$ (long dashed lines). Only $(i)$ is sensitive to the asymmetry parameter.}
\label{fig:qnrj}
\end{figure}

The resilience of the cosine-cosine coupling to strong drives lies in the selection rules enforced by the symmetries of the interaction Hamiltonian (section~\ref{sec:averaging}). Specifically, in the ideal case of a symmetric SQUID $d=0$, only transitions involving an even number of drive photons are allowed at the upper sweet spot $\varphidc=0$. In Fig.~\ref{fig:qnrj}~(Left), we test this resilience by plotting the Floquet spectrum of Hamiltonian~(\ref{eq:sc_hamiltonian}) for the first 12 dressed transmon states $\varepsilon_k(n)$ and multiple values of the offset charge. We distinguish two types of states. The low-lying energy states $k<6$ show little charge sensitivity initially, and they are strongly renormalized by the Floquet drive. On the contrary, the states $k\geq6$ -- initially above the Josephson potential well (Fig.~\ref{fig:matelts}) --  are always strongly charge sensitive, but they disperse weakly with the drive. Overall, there is a large number of multi-excitation resonances in this spectrum. Which ones are activated, and are they harmful to the system stability?

We answer this question with a careful inspection of the spectrum for a single value of the offset charge, see Fig.~\ref{fig:qnrj}~(Center). There, we label every state with its replica index, starting with $m$ for the ground state at $n=0$, and gaining $+1$ every time a quasi-energy level falls into the next Brillouin zone. This labeling lets us identify the parity of the multi-excitation resonances. For instance, the one between states $0$ and $6$ involves a single drive photon, which is forbidden by the ideal cosine-cosine coupling. On inset $(i)$ we check that these two quasi-energy levels do not hybridize when $d=0$ (solid lines), and that a small gap opens for finite asymmetry (dashed lines). Yet even for 2\% asymmetry, the gap is only of 200~kHz, most likely due to the smallness of $J_1(\thetaac)$ at this drive amplitude (Fig.~\ref{fig:bessel}). On the contrary, the multi-excitation resonance between states $0$ and $11$ involves two drive photons, which is allowed by the coupling scheme. On inset $(ii)$ we check that these two quasi-energy levels hybridize, regardless of the asymmetry. Yet the gap is only $\sim15$~kHz, which is a consequence of the smallness of the corresponding matrix element (Fig.~\ref{fig:matelts}). 
\begin{figure}[h]
\includegraphics[]{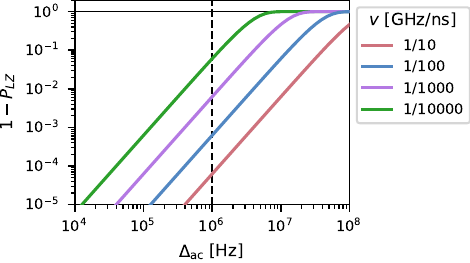}
\caption{\textbf{Landau-Zener passage probability}. The probability of a diabatic transition through a Landau-Zener crossing reads $P_{LZ}=\exp \big(-\pi\Delta_\mathrm{ac}^2/2v\big)$ where $\Delta_\mathrm{ac}$ is the minimal distance between the two energy levels (the gap), and $v$ the speed at which the crossing is traversed. When driving the resonator to the collapse point, Landau-Zener crossings are traversed from a few GHz distance to resonance in a characteristic time $\tau$ set by the pulse sequence. For a square excitation, $\tau$ is lower-bounded by the resonator characteristic ring-up time $1/\kappa$ (51~ns in our device). With a three-segment pulse (section~\ref{sec:RO_opt}), the ring-up time $\tau$ can be significantly reduced. Overall, in our parameter regime, gaps smaller than 1~MHz are traversed diabatically.}
\label{fig:lz}
\end{figure}

In the end, all the resonances involving one of the transmon computational states are either protected by the coupling symmetry, or their activation is heavily suppressed in our parameter regime. Even though finite SQUID asymmetry lifts protection in realistic devices, the resulting gaps remain well below 1~MHz. In Fig.~\ref{fig:lz}, we check that given our typical ramp-up speeds ($v>10^{-3}$~GHz/ns), those gaps are traversed diabatically and lead to negligible Landau-Zener mixing. Together with this highly regular Floquet spectrum, our ability to drive large occupations in the resonator in a short timescale gives the system its resilience to multi-excitation resonances.

\subsubsection{Agreement with averaged theory}
\begin{figure}[t]
\includegraphics[]{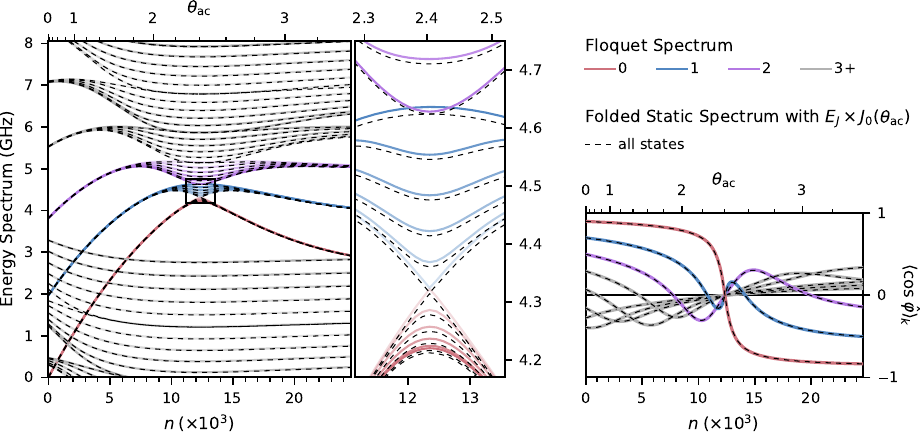}
\caption{\textbf{Agreement with RWA}. Floquet analysis of Hamiltonian~(\ref{eq:sc_hamiltonian}) and diagonalization of Hamiltonian~(\ref{eq:H_fluxdriven_sym_rwa}) computed with $E_C/h=103$~MHz, $E_J/h=5.22$~GHz, $\omega_d/2\pi=8.054$~GHz, $d=0$, $\varphidc=0$ and $p\varphi_{zpf}=0.0216$. (Left) Floquet spectrum (full lines, offset charge from 0 to 0.5 dark to light colors) and static spectrum modulo $\omega_d$ (black dashed lines) of the first 8 levels versus drive strength quantified by the occupation of the resonator (bottom x-axis) and the equivalent flux-drive amplitude (top x-axis). The inset shows a close-up on the collapse region as marked by the black rectangle on the wide spectrum. (Right) Expectation values of $\cos\hat{\varphi}$ for the dressed transmon states computed with the Floquet and static theories.}  
\label{fig:agreement}
\end{figure}

The absence of strongly activated multi-excitation resonances in the Floquet spectrum is the key requirement for the application of the rotating-wave approximation described in section~\ref{sec:averaging}. There, it was shown that the flux-driven transmon can be accurately described by the Hamiltonian of a static transmon with a renormalized Josephson energy $E_J\times J_0(\thetaac)$. In Fig.~\ref{fig:agreement}, we check this assertion by comparing the spectra and state representations of the complete and approximate theories.

First, we compute the eigenstates $\ket{k(n,N_g)}$ and eigenenergies $E_k(n,N_g)$ of the averaged theory. To compare the static spectrum with the Floquet one, we plot the former modulo $\omega_d$ to mimick the single Brillouin zone representation of the latter. In doing so, they line-up almost perfectly (Fig.~\ref{fig:agreement}~left). Only near the collapse they deviate, as evidenced by the presence of a gap in the Floquet spectrum between the first and second levels at the collapse when $N_g=0$, while the static theory predicts a degeneracy. It appears that this feature is entirely captured by the second-order correction to the averaged Hamiltonian. It amounts to an added term $E_{J2}(\thetaac)\cos2\hat{\varphi}$ with \cite{Venkatraman2022, Wang2024}:
\begin{equation}
    E_{J2}(\thetaac) = \frac{E_CE_J^2}{(\hbar\omega_d)^2} \sum_{n>0} \Bigg(\frac{J_{2n}(\thetaac)}{n}\Bigg)^2\;.
\end{equation}
This $\cos2\hat{\varphi}$ term is the quantum analog of the Kapitza contribution to the dynamical stabilization of the inverted pendulum \cite{Butikov2001}. It is a distinct mechanism from the inversion of the Josephson potential presented in the main text, though it can also stabilize inverted transmon states. Crucially, it can also be understood as mediating the tunneling of pairs of Cooper pairs, a highly sought-after resource for the design of fully-protected qubits \cite{Smith2020, Roverch2026, Feldstein-Bofill2026, Zhurbina2026}. In our system, we estimate at the collapse $E_{J2}/h=17$~MHz which is much less than the charging energy, placing the circuit away from any protected regime.

Second, to compare the dressed transmon states of both theories, we look at their charge-averaged expectation values on $\cos\hat{\varphi}$:
\begin{alignat}{2}
    &\text{Static:} \quad &&\langle\cos\hat{\varphi}\rangle_k=\int_0^1 \mathrm{d}N_g \bra{k(n,N_g)} \cos\hat{\varphi} \ket{k(n,N_g)}\;, \\
    &\text{Floquet:} \quad &&\langle\cos\hat{\varphi}\rangle_k=\int_0^1 \mathrm{d}N_g\:\frac{1}{T}\int_0^T \mathrm{d}t \bra{k(n,N_g;t)} \cos\hat{\varphi} \ket{k(n,N_g;t)}\;.
\end{alignat}
The static theory reproduces exactly the evolution of the Floquet matrix elements with the drive amplitude (Fig.~\ref{fig:agreement}~right). Specifically, they change sign at the collapse point, and their ordering flips. In the following, we show that this can be interpreted as a $\pi$-shift of the phase localization of the transmon states. Let $\ket{0}$ and $\ket{\pi}$ be two states of the transmon Hilbert space such that their wavefunctions in phase representation $\psi_{0,\pi}(\varphi)=\bra{\varphi}0,\pi\rangle$ satisfy the relation: $\psi_{\pi}(\varphi)=\psi_{0}(\varphi+\pi)$. Further assuming that they have a definite parity $\psi_{0,\pi}(-\varphi)=(-1)^p \psi_{0,\pi}(\varphi)$ with $p\in\{0,1\}$, we can write:
\begin{equation}
    \bra{0} \cos\hat{\varphi} \ket{0}
    \coloneq \int_{\text{-}\pi}^\pi \mathrm{d}\varphi |\psi_{0}(\varphi)|^2\cos\varphi 
    \underset{\varphi\:\rightarrow\:\pi-\varphi}{=} - \int_0^{2\pi} \mathrm{d}\varphi |\psi_{\pi}(\varphi)|^2\cos\varphi
    \eqcolon -\bra{\pi}
    \cos\hat{\varphi}
    \ket{\pi}\;.
    \label{eq:matel_shift}
\end{equation}
Knowing that the wavefunctions of a transmon are either even or odd in $\varphi$, we can use Eq.~(\ref{eq:matel_shift}) to interpret the change of sign of the expectation values of $\cos\hat{\varphi}$ at the collapse as a shift of the state localization by $\pi$. One should notice that the parity of the wavefunctions is also well-defined for a flux-tunable transmon biased at zero flux, regardless of the SQUID asymmetry. As a consequence this interpretation is robust to fabrication imperfection.

\subsection{Readout through Cosine-Cosine Coupling}
\label{sec:readout}
In this part we will show how the cosine-cosine coupling can be leveraged for transmon readout, and reveal the localization of the transmon states.
In the previous part we neglected the quantum fluctuations of the resonator in the presence of the strong drive, and looked at the transmon renormalization only. In the following, we consider the full quantum system in the limit $\varphidc=0$ and $d=0$, and focus on the resonator properties conditioned on the dressed transmon states. This is most conveniently done in a frame displaced around the unconditional resonator field $\alpha(t)=\sqrt{n}\sin\omega_d t$, and rotating at a frequency $\omega_d$ for the $\ha$-mode. There, under the rotating-wave approximation, Hamiltonian~(\ref{eq:full_hamiltonian}) is transformed up to second order in $(p\varphi_{zpf})^2$ into:
\begin{equation}
\begin{split}
    \hat{\mathcal{H}} =\: &\hbar(\omegaa-\omega_d) \had\ha+ 4E_{C}\big( \hat{N} - N_g\big)^2 - \tilde{E}_{J}(\thetaac) \cos\hat{\varphi} \\ 
    &+ \hbar Z_0(\thetaac)\: \had\ha \cos\hat{\varphi} + \hbar Z_1(\thetaac) \big(\ha + \had \big) \cos\hat{\varphi} + \hbar Z_2(\thetaac) \big(\ha^2 + \hat{\mathrm{a}}^{\dagger 2} \big) \cos\hat{\varphi}\;,
\end{split}
\end{equation}
where we remind that $\thetaac=p\varphi_{zpf}\sqrt{n}$, and we find:
\begin{align}
    \tilde{E}_J(\thetaac) &=  \big(1 - \tfrac{1}{8}p^2\varphi_{zpf}^2\big)E_{J} J_0(\thetaac)\;, \\
    \hbar Z_0(\thetaac) &= \frac{1}{4} (p\varphi_{zpf})^2 E_{J} J_0(\thetaac)\;, \\
    \hbar Z_1(\thetaac) &= \frac{1}{2} p\varphi_{zpf} E_{J} J_1(\thetaac)\;, \\
    \hbar Z_2(\thetaac) &= \frac{1}{8} (p\varphi_{zpf})^2 E_{J} J_2(\thetaac)\;.
\end{align}
The first term displays the renormalization of the Josephson energy imparted by the resonator phase fluctuations, akin to the renowned Lamb shift. The following three correspond to the different readout channels: the dispersive channel coupling to $\had\ha$, the displacement channel coupling to $\ha+\had$, and the squeezing channel coupling to $\ha^2 + \hat{\mathrm{a}}^{\dagger 2}$.

\subsubsection{Dispersive channel}
The dispersive channel with amplitude $Z_0(\thetaac)$ is a \emph{native} one: since $J_0(0)=1$, it is opened even in the absence of the resonator drive. It is however closed at the collapse. It yields a resonator frequency shift conditioned on the transmon state:
\begin{equation}
    \chi_{k}(\thetaac, N_g) = Z_0(\thetaac) \times \bra{k(\thetaac, N_g)} \cos\hat{\varphi} \ket{k(\thetaac, N_g)}\;.
\end{equation}
This dispersive interaction is also known as cross-Kerr. Indeed, introducing the bosonic representation of the transmon circuit $\hat{\varphi} = \varphi_t (\hat{\mathrm{b}}+ \hat{\mathrm{b}}^\dagger)$ with $\varphi_t=(2E_C/|\tilde{E}_J(\thetaac)|)^{1/4}$ and $\hat{b}$ satisfying $[\hat{b}, \hat{b}^\dagger]=1$, the interaction typically reads
$\chi_z(\thetaac,N_g)\had\ha\hat{\mathrm{b}}^\dagger\hat{\mathrm{b}}$, where $\chi_z(\thetaac,N_g) = \chi_{1}(\thetaac,N_g)-\chi_{0}(\thetaac,N_g)$ is the dispersive shift to the qubit transition. This cross-Kerr is said to be \emph{non-perturbative} because it is directly entailed in the cosine-cosine coupling rather than the product of the perturbative expansion of a dipolar coupling mechanism. It is also said to be \emph{longitudinal}, because the cosine-cosine interaction couples to the $Z$-axis of the qubit transition.

In the generalized transmon limit $|\tilde{E}_J(\thetaac)|\gg E_C$, charge sensitivity is suppressed and one finds:
\begin{equation}
    \chi_z(\thetaac) = -\frac{1}{8}(p\varphi_{zf})^2 \sqrt{8E_CE_J|J_0(\thetaac)|}\;.
\end{equation}
One should notice that in this expression the sign change of both the channel amplitude and the $\cos\hat{\varphi}$ expectation values conspire in the inverted regime to yield an always-negative dispersive shift. This fulfills the intuition that this channel is the byproduct of the transmon anharmonicity, which is always negative whether the transmon is inverted or not.

\subsubsection{Displacement channel}
The displacement channel with amplitude $Z_1(\thetaac)$ is closed in the absence of the resonator drive, but it is the dominant one around the collapse. It yields a drive term conditioned on the transmon state, with amplitude:
\begin{equation}
    \zeta_{k}(\thetaac, N_g) = Z_1(\thetaac) \times \bra{k(\thetaac, N_g)} \cos\hat{\varphi} \ket{k(\thetaac, N_g)}\,.
\end{equation}
When the unconditional drive is set on resonance ($\omega_d=\omegaa)$, this added term displaces the resonator field in a direction orthogonal to its unconditional growth, and with a conditional amplitude:
\begin{equation}
    \delta \alpha_{\ket{k}}(\thetaac, N_g) = p \varphi_{zpf} \frac{E_J}{\hbar\kappa} J_1(\thetaac) \times \bra{k(\thetaac, N_g)} \cos\hat{\varphi} \ket{k(\thetaac, N_g)}\;.
    \label{eq:condtional_disp}
\end{equation}
Since $J_1$ is always positive around the collapse, the direction of this conditional displacement translates the sign of the expectation values of $\cos\hat{\varphi}$, hence the phase localization of the transmon states (Eq.~(\ref{eq:matel_shift})).

This displacement channel arises from the native dispersive one when viewed in the displaced frame of the resonator. Indeed, just like a transmon-resonator system dispersively coupled through $\chi_z \had\ha \hat{\mathrm{b}}^\dagger\hat{\mathrm{b}}$ yields in the displaced frame $\ha\rightarrow \ha+\sqrt{n}$ the conditional drive $\zeta_z (\ha+\had) \hat{\mathrm{b}}^\dagger \hat{\mathrm{b}}$ with $\mathrm{d}\zeta_z / \mathrm{d}\sqrt{n} = \chi_z$, here we find:
\begin{equation}
    \frac{\mathrm{d}Z_1(\thetaac)}{\mathrm{d}\sqrt{n}} = Z_0(\thetaac) \times \bigg( 1 - \frac{J_2(\thetaac)}{J_0(\thetaac)} \bigg) \xrightarrow[n\rightarrow0]{} Z_0(\thetaac)\;.
\end{equation}
This approach has been known in the literature as a way to reproduce the \emph{longitudinal readout} paradigm first introduced by Didier and co-authors \cite{Didier2015}. It is applicable whether the underlying cross-Kerr is native and coming from a \emph{longitudinal coupling} mechanism like in this paper or in Refs.~\cite{Chapple2025, Mori2025b, Salunkhe2025, Hazra2025}, or not \cite{Touzard2019, Ikonen2019}.

\subsubsection{Squeezing channel}
The squeezing channel $Z_2(\thetaac)$ opens for finite drive amplitudes, but owing to the smallness of $p\varphi_{zpf}$ it is greatly suppressed compared to the displacement one. We will neglect it in the rest of the analysis.

\newpage
\section{Observation of Collapse and Inversion through Readout}
\subsection{Analysis of the Readout Experiment}
\label{sec:SI_AnalysisRO}

\begin{figure}[b]
\includegraphics{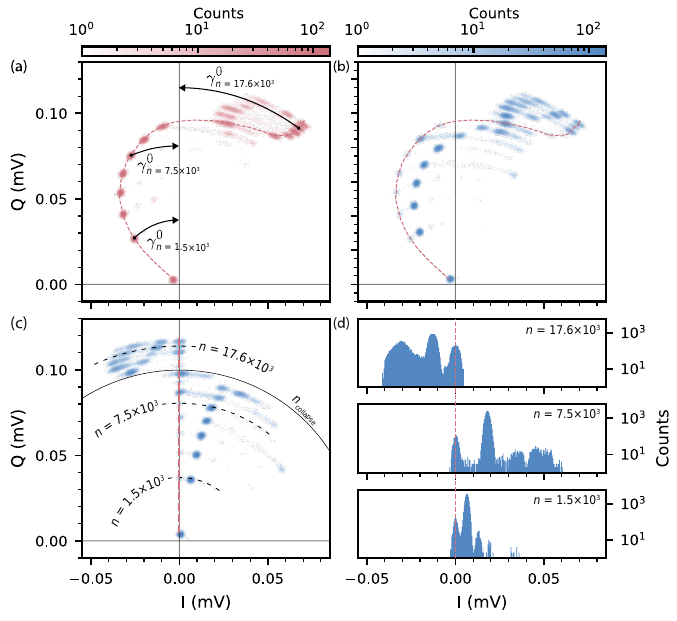}
\caption{\label{fig:figSI_AnalysisBC} \textbf{Analysis of the readout experiment.}
Unprocessed IQ histograms for preparation of the transmon in $|0\rangle$ (a) and $|1\rangle$ (b) for a range of steady-state resonator photon numbers, converted from readout power using the method described in section \ref{sec:photon_number_cal}. The dashed salmon-colored line passes through the maxima of the $|0\rangle$-distributions, determined after Gaussian smoothing of the unprocessed IQ histograms for preparation in $|0\rangle$. $\gamma_n^0$ are the photon-number dependent angles from the $|0\rangle$-distribution maxima to the y-axis. (c) IQ histograms for preparation in $|1\rangle$ rotated by $\gamma_n^0$. This photon-number dependent rotation ensures centering of the dressed ground states along the line $I = 0$~mV, as indicated by the dashed salmon-colored line. (d) Histograms of the I quadrature for photon numbers $n = 1.5\times10^3$, $7.5\times10^3$ and $17.6\times10^3$.}  
\end{figure}

In this section, we detail the analysis procedure used to construct the I quadrature histogram of Fig.~4(b) in the main text. As previously described, the cosine-cosine transmon-resonator coupling realized by our circuit design implements a non-perturbative cross-Kerr interaction similar to refs. \cite{Dassonneville2020,Chapple2025,Mori2025a,Mori2025b,Salunkhe2025,Hazra2025,Hazra2026}. As in standard dispersive readout, information about the state of the transmon is encoded in the amplitude and phase, or equivalently the in-phase ($I$) and quadrature ($Q$) components, of a near-resonant tone reflected from the resonator \cite{Blais2021}. Repeated application of the same readout tone leads to a distribution in the IQ plane of the resonator, corresponding to a probability distribution of the transmon occupation. In Fig.~\ref{fig:figSI_AnalysisBC}(a,b), we show these distributions for preparation of the transmon in $|0 \rangle$ (a) and $|1\rangle$ (b) for a range of resonator photon numbers. A three-segment readout pulse (see section~\ref{sec:RO_opt}) with fixed frequency $\omega_d/2\pi = 8.05363$~GHz and a length of $2200$~ns is used. To probe the transmon states at a steady-state resonator photon number, the readout pulse is integrated in the interval $[1400,2000]$~ns. With increasing readout power, the measured IQ distributions for preparation in both $|0\rangle$ and $|1\rangle$ increase in distance from the origin. In addition, they experience a power-dependent phase shift. As we will see in section \ref{sec:SI_Sim_IntraField}, the resonator nonlinearity alone is not sufficient to explain this rotation, leaving us to suspect the nonlinearity of the room-temperature hardware at high power. Nevertheless, the phase shift does not impact the analysis, since the distributions for $\ket{0}$ and $\ket{1}$ are affected in the same way.\\
\indent To reduce the dimensionality of this dataset and facilitate a simpler interpretation, we apply the following analysis steps. Let us first focus on preparation in $|0\rangle$. At every readout power, the location of the IQ distribution's maximum is determined. To reduce noise in the location of the maximum, a Gaussian smoothing of the unprocessed histogram is performed prior. We obtain the dashed salmon-colored line in Fig.~\ref{fig:figSI_AnalysisBC}(a,b), which at every power passes through the centroid of the highest density transmon state in the $|0\rangle$-distribution. We therefore identify this line with the dressed ground state of the driven system. We determine the photon-number dependent angle from the dressed ground state to the y-axis, $\gamma_n^0$. Next, the IQ distributions for preparation in $|1\rangle$ are rotated by $\gamma_n^0$, yielding Fig.~\ref{fig:figSI_AnalysisBC}(c). These rotated IQ distributions now show a clear inversion about $I = 0.0$~mV at $n_{\textrm{collapse}}$. The fact that all counts not identified as ground move from $I>0$~mV to $I<0$~mV, shows that our determination of the dressed-ground-state reference was appropriate. In the final step of the analysis, the rotated IQ distributions are projected onto the I quadrature, resulting in one-dimensional histograms. Examples for $n = 1.5\times10^3$, $7.5\times10^3$ and $17.6\times10^3$ are shown in Fig.~\ref{fig:figSI_AnalysisBC}(d). These one-dimensional histograms are plotted against steady-state photon number to produce Fig.~4(b) of the main text. It is important to note that the chosen I quadrature projection reduces the SNR between the transmon states, which lie on circle arcs with radii set by the readout power (see Fig.~\ref{fig:figSI_AnalysisBC}(c)). However, since the purpose of this measurement was to elucidate the features of the collapse and inversion, we opted for a conceptually simple projection that fulfills this purpose.

\subsection{Simulation of the Intracavity Field}
\label{sec:SI_Sim_IntraField}
\begin{figure}[b]
\includegraphics{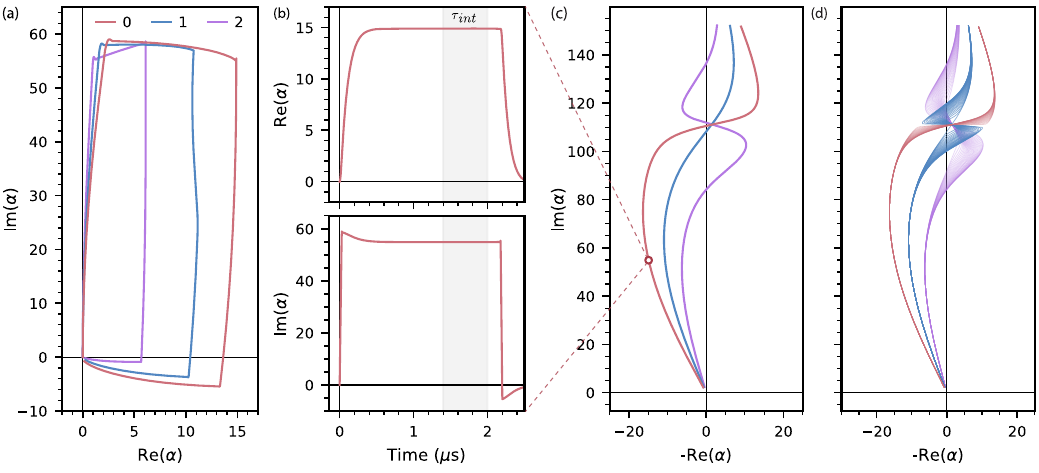}
\caption{\label{fig:figSI_ResFieldSim} \textbf{Simulation of the intracavity field.} (a) Evolution of the intracavity field in phase space under the application of a readout pulse for initial transmon states $|0\rangle$, $|1\rangle$, and $|2\rangle$. (b) Real and imaginary parts of the simulated resonator field for initial state $|0\rangle$ as a function of time. The field is integrated over the shaded region of length $\tau_{int} = 600$~ns, corresponding to the integration window used in the experiment, leading to a point in phase space marked by the open circle in (c). The phase space trajectories in (c) show the integrated intracavity fields as a function of readout power, evaluated at a single offset charge value $N_g = 0.0$. These trajectories represent how the dressed transmon states
evolve in the phase space of the resonator with increasing steady-state resonator occupation. Note that we plot against $-\textrm{Re}(\alpha)$, see text for a discussion. (d) Same simulation of the integrated field repeated for a range of offset charges in the interval [0,0.5].}  
\end{figure}
To explain the measured evolution of the dressed transmon states along the I quadrature of the resonator, we simulate the intracavity field of the resonator (see Fig.~4(c) of the main text) using the quantum Langevin equation \cite{Blais2021}: 
\begin{equation}
\dot{\ha} = \frac{i}{\hbar}\left [\hat{\mathcal{H}},\ha \right]-\frac{\kappa}{2}\ha+\sqrt{\kappa}\ha_{\textrm{in}}(t).
\label{eq:SI_qLangevin}
\end{equation}
Here, we have made the approximation $\kappa \approx \kappa_{\textrm{ext}}$, which is appropriate for our device parameters. The Hamiltonian of Eq.~(\ref{eq:full_hamiltonian}) at the operation point $\varphi_{dc} = 0$ and assuming $d=0$ reduces to
\begin{equation}
\hat{\mathcal{H}} = \hbar \omegaa \had\ha+4E_C\left (\hat{N}-N_g \right )^2-E_J\cos \frac{p\varphi_{zpf}(\ha+\had)}{2} \cos \hat{\varphi},
\label{eq:SI_H_Langevin}
\end{equation}
where we have dropped the $\Sigma$ subscript of the transmon mode. We evaluate $[\hat{\mathcal{H}},\hat{\mathrm{a}}]$ by expressing $\cos \left [ p\varphi_{zpf}(\ha+\had)/2 \right ] = [\hat{D}(ip\varphi_{zpf}/2) + \hat{D}(-ip\varphi_{zpf}/2)]/2$ and making use of the commutation relation $[\hat{D}(\alpha),\hat{\mathrm{a}}] = -\alpha \hat{D}(\alpha)$. After transforming to a frame rotating at the frequency of the input field $\omega_d$, i.e. $\hat{\mathrm{a}} \rightarrow \hat{\mathrm{a}}\exp(-i\omega_d t)$, we find
\begin{equation}
\dot{\ha} = - \bigg(\frac{\kappa}{2} + i\Delta\bigg)\ha - i\frac{p\varphi_{zpf}}{2} \frac{E_J}{\hbar} \cos \hat{\varphi} \sin \bigg( \frac{p\varphi_{zpf}}{2} \left [\ha e^{-i\omega_dt}+\had e^{+i\omega_dt} \right ] \bigg) e^{+i\omega_dt}+\sqrt{\kappa}\ha_{\textrm{in}}(t)e^{+i\omega_d t},
\label{eq:SI_qLangevin2}
\end{equation}
where $\Delta = \omegaa-\omega_d$. The semiclassical equation of motion is found by taking the expectation value of Eq. (\ref{eq:SI_qLangevin2}) with the state $|k(n, N_g)\rangle \otimes |\alpha\rangle$, where $|k(n, N_g)\rangle$ are the dressed transmon states that depend on the intracavity photon number $n$ and offset charge $N_g$. Neglecting quantum fluctuations of the input field, this yields
\begin{equation}
\dot{\alpha} = - \bigg(\frac{\kappa}{2} + i\Delta \bigg) \alpha -i\frac{p\varphi_{zpf}}{2} \frac{E_J}{\hbar} \langle \cos \hat{\varphi} \rangle_k(n,N_g) \times e^{-\frac{1}{2}(p\varphi_{zpf}/2)^2} \sin \Big( p\varphi_{zpf} \textrm{Re}\left [\alpha e^{-i\omega_d t} \right ] \Big) e^{+i\omega_d t} + \sqrt{\kappa} \alpha_{\textrm{in}}(t)e^{+i\omega_dt}\;,
\label{eq:SI_qLangevinSC}
\end{equation}
where $\langle \cos \hat{\varphi} \rangle_k(n,N_g) = \langle k(n,N_g) |\cos \hat{\varphi}\,| k(n,N_g) \rangle$.
We specify the classical input field as $\alpha_{\textrm{in}}(t) = \varepsilon_d(t)/\sqrt{\kappa} \cos \omega_dt$, where $\varepsilon_d(t)$ matches the three-segment pulse envelope used in the experiment (section~\ref{sec:RO_opt}). Finally, to further simplify this equation of motion, we apply the Jacobi-Anger expansion to $\sin  (p\varphi_{zpf} \textrm{Re} [\alpha \exp(-i\omega_dt)])$ and perform the RWA to neglect any fast rotating terms: 
\begin{equation}
\dot{\alpha} = - \bigg(\frac{\kappa}{2} +i\Delta \bigg)\alpha + \frac{\kappa}{2} \bigg( \frac{\varepsilon_d(t)}{\kappa} +  \delta\alpha_{\ket{k}}(n,N_g) \times e^{-\frac{1}{2}(p\varphi_{zpf}/2)^2} e^{i(\Theta-\pi/2)}  \bigg)\;,
\label{eq:SI_qLangevinSC_RWA}
\end{equation}
where $\Theta = \arg(\alpha)$. The second term displays the unconditional and conditional displacements. The latter differs from Eq.~\ref{eq:condtional_disp} by a small renormalization imparted by the quantum fluctuations of the resonator internal field, that was omitted at the order of the previous calculation. The complex phase factor shows that the conditional displacement is orthogonal to the unconditional one. This derivation elucidates the fact that the cosine-cosine coupling between transmon and resonator allows us to probe these expectation values through a measurement of the resonator field.\\
\indent Eq. (\ref{eq:SI_qLangevinSC_RWA}) is numerically integrated for a given input pulse $\varepsilon_d(t)$. All system parameters that enter the equation are determined from independent measurements. The dressed transmon states $|k(n,N_g) \rangle$, required for the calculation of the expectation values of $\cos \hat{\varphi}$, are obtained from Floquet analysis, see section \ref{sec:SI_FloquetAnalysis}. Interpolation is used to include the expectation values as a general $\alpha$-dependent coefficient in the differential equation. Fig.~\ref{fig:figSI_ResFieldSim}(a) shows the evolution of the intracavity field in phase space for a certain $\varepsilon_d(t)$. The equation of motion is solved for initial states $|0\rangle$, $|1\rangle$, and $|2\rangle$. Fig.~\ref{fig:figSI_ResFieldSim}(b) shows the real and imaginary part of the trajectory of $|0\rangle$ as a function of time. To match the analysis of the experimental data, the resulting intracavity field is integrated over a window from $t=1400$~ns to $t=2000$~ns. It is clear that the intracavity field has reached steady-state in this interval. This allows us to identify the integrated field, represented by the open circle in Fig.~\ref{fig:figSI_ResFieldSim}(c), with a steady-state resonator occupation. This procedure is repeated for input pulses with increasing amplitude, leading to trajectories as a function of photon number in (c). These trajectories are evaluated at a single offset charge, $N_g=0.0$, i.e. the dressed transmon states are calculated for this offset charge. Repeating the simulation of the trajectories for offset charge values between 0.0 and 0.5 yields the trajectories in Fig.~\ref{fig:figSI_ResFieldSim}(d). Finally, to allow for a comparison of the simulated intracavity field and the measured output field, we apply a rotation of the trajectories based on the angle of the trajectory of the dressed ground state (Fig.~4(c) of the main text). As in the analysis of the measured output field (section~\ref{sec:SI_AnalysisRO}), this rotation removes any global phase shifts and reveals the relative difference between the states. After referencing the trajectories to the dressed ground state, only a scalar multiplication is required to match the simulated trajectories to the measured ones. We note that there is a sign difference between data and theory, which prompted us to plot $-\mathrm{Re}(\alpha)$ in Fig.~\ref{fig:figSI_ResFieldSim}(c,d). Since the simulation correctly predicts the measured resonator frequency shift with power, we suspect it to be a trivial mirroring that occurs during demodulation of the data. Nonetheless, it does not change the relative difference between the trajectories.

\subsection{Additional Features in the Readout Experiment}
The dynamical renormalization of the potential further manifests itself in other signatures that we observe in the data of Fig.~4 of the main text. The broadening of the dressed excited state signal around the collapse point can be attributed to the restoration of charge sensitivity, which was also inferred from Fig.~3 of the main text. Beyond that, the readout experiment enables us to observe an increase in the population of the higher excited states from $n > 8\times10^3$. Before the collapse, this is expected to be a result of thermal excitation as the qubit frequency decreases and its energy becomes comparable to the thermal energy. From the collapse point onward, the energy spectrum becomes gapless during the measurement pulse when $N_g = 0$ or $N_g=0.5$. As the offset charge is not controlled in the experiment, its random sampling at constant $n$ causes additional population transfer from the dressed excited state through Landau-Zener transitions \cite{Grifoni1998,Ivakhnenko2023}. This effect is studied more quantitatively in section~\ref{sec:SI_DUST}.

\section{Calibration \& Optimization}
\subsection{Readout Mechanism \& Optimization}
\label{sec:RO_opt}

In this section, we present the details of the readout mechanism and its optimization, a key component for state preparation and measurement in Fig.~3 and 4 of the main text. As described in ref. \cite{Chapple2025} and also detailed in section \ref{sec:readout}, a cosine-cosine coupling between transmon and resonator implements an effective longitudinal readout after displacement of the resonator field by a microwave drive. Through our coupling, this drive is converted to a flux modulation of the transmon at the frequency of the resonator, allowing us to conceptually understand this as the protocol described in ref. \cite{Didier2015}. In the limit of large displacement, attainable through the robustness of this coupling scheme to multi-excitation resonances (see section \ref{sec:SI_FloquetAnalysis}), the conditional drive term dominates the cross-Kerr term, thus implementing an effective longitudinal readout.\\
\begin{figure}[b]
\includegraphics{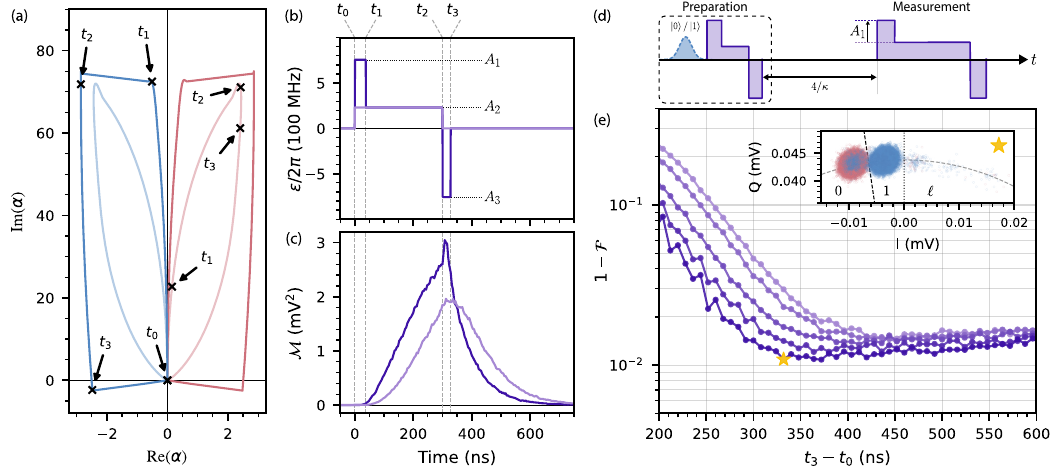}
\caption{\label{fig:figSI_ReadoutOpt} \textbf{Readout mechanism \& optimization.} (a) Intracavity phase space trajectories for transmon in $|0\rangle$ (salmon) and $|1\rangle$ (blue) under the application of a three-segment (dark color) and rectangular (light color) pulse. The pulse shapes are shown in (b). The edges of the three-segment pulse are indicated by the time labels $t_i$. These times are marked by crosses on the phase space trajectories in (a). (c) Measurement rate as a function of time for the pulse shapes in (b). (d) Pulse sequence used to measure the three-state assignment fidelity $\mathcal{F}$. The transmon is prepared in either $|0\rangle$ or $|1\rangle$ and subsequently measured by a three-segment readout pulse of which the total length, $t_3-t_0$, and the amplitude of the first segment, $A_1$, are varied. (e) Assignment infidelity, $1-\mathcal{F}$, as a function of pulse length. $A_2$ is fixed to an equivalent photon number of $n \approx 5\times10^3$. $A_1$ is varied from $A_2$ (light) to $3.4A_2$ (dark). We find an optimal fidelity of 98.9\% in 332~ns, marked by the star. The inset shows the distribution of the $5 \times 10^4$ readout events of the optimal pulse in the IQ plane. The dashed line is the discrimination threshold between the computational states (outcomes $0$ and $1$), the dotted one between the excited and the leakage states, labeled $\ell$.}  
\end{figure}
\indent However, in contrast to direct flux modulation, which instantaneously activates the conditional displacement, a microwave drive near resonance will only displace the resonator on a timescale of $1/\kappa$. To efficiently activate it, a rapid displacement of the resonator field is required through appropriate pulse shaping. Here, we exemplify this by comparing a standard rectangular pulse to a three-segment pulse, see Fig.~\ref{fig:figSI_ReadoutOpt}(b). The amplitudes $A_1$ and $A_3$ of the three-segment pulse are set to the maximum output value of the control hardware. The amplitude of the second segment $A_2$ determines the target steady-state photon number. The lengths of the first and third segments, which displace and deplete the resonator to and from the steady state, depend on the ratios $A_1/A_2$ and $A_3/A_2$, respectively \cite{Chatterjee2025}. We simulate the intracavity field of the resonator for transmon states $|0\rangle$ and $|1\rangle$ and plot the phase space trajectories for both pulses in Fig.~\ref{fig:figSI_ReadoutOpt}(a). The rectangular pulse leads to the familiar evolution of the states in dispersive readout \cite{Blais2021}, where the separation between $|0\rangle$ and $|1\rangle$ grows as the photon number in the resonator rings up. A similar trajectory back to the origin is followed after $t_2$ when the rectangular pulse ends. In contrast, the first segment of the three-segment pulse first rapidly displaces the resonator to a high photon number. During the second segment, the activated conditional displacement drives a linear separation between the states. Finally, the third segment depletes a large fraction of the resonator photons, leaving only a small fraction to decay exponentially. \\
\\
\indent In Fig.~\ref{fig:figSI_ReadoutOpt}(c), we investigate the difference between the rectangular and three-segment pulse experimentally by determining the measurement rate, $\mathcal{M}(t) \propto |\alpha^0_{\textrm{out}}(t)-\alpha^1_{\textrm{out}}(t)|^2$, with $\alpha^{0,1}_{\textrm{out}}(t)$ the average instantaneous output fields of the resonator for the transmon prepared in $|0\rangle$ and $|1\rangle$. Here we observe that after a time $t_1$, the duration of the first segment, the measurement rate increases significantly faster for the three-segment pulse than for the rectangular one. This signifies that the linear conditional displacement, activated by the first segment, leads to a larger separation of the states at shorter times. This difference is further amplified by our choice of a 300~ns readout pulse, which for our $\kappa/2\pi = 3.1$~MHz is shorter than $10/\kappa$. At $t_2$, this leads to a significant difference in measurement rate between the pulses. After $t_2$, the separation for the rectangular pulse follows the natural ring-down of the resonator. The third segment of the three-segment pulse quickly depletes the resonator, speeding up the decay of $\mathcal{M}$ by $\sim$25\%. We emphasize that we not only make use of this three-segment pulse to improve the signal-to-noise ratio (SNR) of the readout, but also to minimize leakage to non-computational states by diabatically traversing activated multi-excitation resonances (see Fig.~\ref{fig:lz}). Finally, it should be noted that our pulse bears resemblance to the CLEAR pulse \cite{McClure2016} and the four-segment pulses used in refs. \cite{Chatterjee2025,Hazra2025}. Due to our $\chi_z/\kappa$ ratio of $1/25$, an additional fourth segment did not lead to a further improvement of the resonator depletion time.\\
\\
The comparison of the measurement rates shows that appropriate pulse shaping activates an effective longitudinal readout and improves the SNR at fixed pulse duration. We further explore this quantitatively by measuring the three-state assignment fidelity $\mathcal{F}$ as a function of total pulse length for various values of $A_1$. It is defined as $\mathcal{F} = [P(0|0)+P(1|1)]/2$, where $P(i_m|j_s)$ is the probability to measure outcome $i_m$ given prepared transmon state $j_s$. As can be seen in the inset of Fig.~\ref{fig:figSI_ReadoutOpt}(e), our readout distinguishes between the computational states and all leakage states such that $i_m \in \{0,1,\ell\}$. Due to this ability to unambiguously assign the leakage states, $\mathcal{F}$ is a good approximation of the QND fidelity \cite{Hazra2025}, which quantifies the probability that a measurement assigns the correct outcome without affecting the qubit state.\\
\indent Fig.~\ref{fig:figSI_ReadoutOpt}(d) shows the pulse sequence used to measure the assignment fidelity. Here, and in Fig.~4 of the main text, we start with a state preparation that consists of an optional $\pi$ pulse followed by a readout pulse. Due to our low qubit frequency of 1.972 GHz, we typically measure a sizable thermal population in $|1\rangle$ of 5-10\%, corresponding to an effective temperature in the range of $30- 40$~mK. The preparation pulse is used to post-select the outcomes based on the application of the optional $\pi$ pulse (retain outcome $1$ if applied, $0$ if not), which reduces the excited state thermal population to a value below 0.1\%. The preparation is followed by the application of the measurement pulse. The amplitude of the first segment is varied, while its length is fixed to 36~ns (corresponding to the maximum ratio $A_1/A_2$ reached in this experiment). This allows us to progressively transform the pulse from a rectangular pulse at $A_1 = A_2$ to a strong three-segment pulse with a maximum of $A_1 = 3.4A_2$. The total length of the pulse is varied by changing the length of the second segment. It is important to note that the signal is amplified by an SPA operated in phase-preserving mode before passing through the rest of the measurement chain. The pulse is then integrated over its full length using a window that optimizes the SNR, $\mathcal{W}(t) \propto |\langle \alpha^0_{\textrm{out}} - \alpha^1_{\textrm{out}}\rangle|$ \cite{Walter2017,Hazra2025}.\\
\indent At every $A_1$ and pulse length, we measure IQ histograms consisting of $5\times10^4$ shots, see inset of Fig.~\ref{fig:figSI_ReadoutOpt}(e). Through linear discrimination thresholds between $|0 \rangle$, $|1\rangle$ and the higher-excited states, we determine state probabilities and calculate the assignment fidelity, $\mathcal{F}$. We plot 1-$\mathcal{F}$ in Fig.~\ref{fig:figSI_ReadoutOpt}(e) as a function of pulse duration. In line with the measurement-rate result, we see that the assignment infidelity decreases with increasing $A_1$. This is especially noticeable for pulses with a total length shorter than $10/\kappa \approx 500$~ns. Here, pulses with a larger $A_1$ are able to drive separation between the qubit states on a timescale shorter than the natural resonator ring-up time. \\
\indent We find an optimal fidelity of 98.9\% for a pulse duration of 332 ns. This optimum is reached for a steady-state photon number of $5\times10^3$, optimized in a separate measurement. We determine that the overlap of the ground and excited state distributions results in an error of $\epsilon_{o}=0.1\%$. Transitions between the ground and excited states contribute $\epsilon_{t}=0.72\%$. The transition error can be explained by thermalization (qubit decay and excitation) during the pulse sequence. We can estimate the thermalization error as $\epsilon_{th} = \tau_w \gamma_w (2 \bar{n}_{th}+1)/2 + \tau_d \gamma_d (2 \bar{n}_{th}+1)/2$ with $\bar{n}_{th}$ the thermal photon number of the bath, $\tau_w$ the wait time after preparation, and $\tau_d$ the pulse duration. $\gamma_{w,d}$ are the relaxation rates during the wait time and readout pulse, respectively. For $\bar{n}_{th} = 0.125$, corresponding to a thermal excited state population of $10\%$, we find that $\epsilon_{t} = \epsilon_{th}$ if we assume that the driven relaxation rate is twice the undriven one ($1/\gamma_w = 71~\mu\textrm{s}$). This points towards a significant measurement-induced relaxation \cite{Thorbeck2024,Huang2026,Spring2026}. Finally, the remaining error contribution $\epsilon_{\ell} = [P(\ell|0)+P(\ell|1)]/2 = 0.26\%$ stems from leakage to higher excited states, which we explore in more detail in section~\ref{sec:SI_DUST}.

\subsection{Photon Number Calibration}
\label{sec:photon_number_cal}

\begin{figure}[b]
\includegraphics{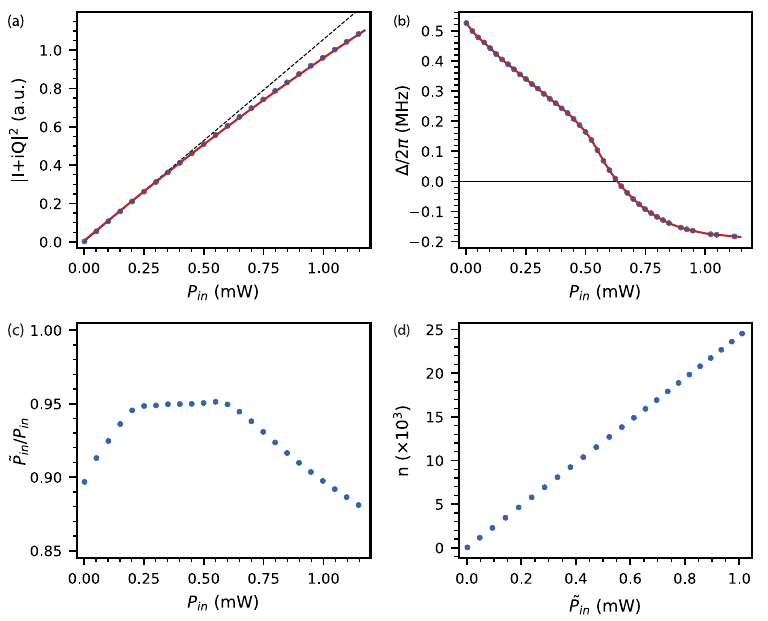}
\caption{\label{fig:figSI_PhotonNumberCal} \textbf{Photon Number Calibration.} (a) Relation between input power  $P_{\textrm{in}}$ set in hardware and measured output power $|I+iQ|^2$. The solid red line is a second-degree polynomial fit. The dashed black line is the expected linear relation between input and output power. (b) Frequency shift of the resonator as a function of $P_{\textrm{in}}$, obtained from a resonator spectroscopy measurement. The resonator frequency is expressed in terms of the detuning $\Delta = \omegaa-\omega_d$ with $\omega_d/2\pi = 8.05363$~GHz. The solid red line is a cubic spline interpolation of the measured data. (c) Power-dependent scaling factor between input power, $P_{\textrm{in}}$, and rescaled power $\tilde{P}_{\textrm{in}} = \nu(P_{\textrm{in}})P_{\textrm{in}}$. Here, $\nu(P_{\textrm{in}})$ combines the contributions from the nonlinear power transfer of our microwave components (a), and from the nonmonotonic frequency shift of the resonator (b). (d) Linear relation between resonator photon number and rescaled input power $\tilde{P}_{\textrm{in}}$, determined from a measurement of the qubit frequency shift.}  
\end{figure}

In the main text, we express the drive power on the resonator in terms of the steady-state photon number. As described there, a fit to the nonlinear qubit frequency shift provides a direct calibration of the power. However, prior to this fitting procedure, we require two additional calibration steps of the input power due to the finite dynamic range of our room-temperature microwave components, and the power-dependent frequency shift of the resonator.\\
\indent Let us first consider the dynamic range of the microwave components. To investigate this, we send an off-resonant tone to the resonator and measure the reflected signal. In Fig.~\ref{fig:figSI_PhotonNumberCal}(a), we plot the measured output power, $|I+iQ|^2$, as a function of the input power $P_{\textrm{in}}$ set in the hardware. The power reflected from the resonator shows a nonlinear dependence on $P_{\textrm{in}}$, indicating that power is transferred nonlinearly through the measurement chain. We confirm that this nonlinear transfer occurs on the input side through a separate measurement of the power (not shown here) before the signal enters the dilution refrigerator. Therefore, a rescaling of the power incident on the resonator is required. The measured relation between input and output power in Fig.~\ref{fig:figSI_PhotonNumberCal}(a) is fitted to a second-degree polynomial $p_2(P_{\textrm{in}})$. Dividing by the expected linear scaling $p_1(P_{\textrm{in}})$ yields the following scaling factor: 
\begin{equation}
\nu_1(P_{\textrm{in}}) = \frac{p_2(P_{\textrm{in}})}{p_1(P_{\textrm{in}})}.
\label{eq:SI_pscale1}
\end{equation}
\indent Next, we consider the effect of the resonator frequency shift on the intracavity power. Fig.~\ref{fig:figSI_PhotonNumberCal}(b) shows the resonator frequency, expressed as $\Delta=\omegaa-\omega_d$, extracted from a resonator spectroscopy as a function of $P_{\textrm{in}}$. In Fig.~3 and 4 of the main text, the resonator is driven at a frequency $\omega_d/2\pi = 8.05363$~GHz. During these measurements, as the input power is increased, the detuning from the drive changes nonlinearly. As a result, the intracavity photon number changes with $\Delta$ according to $n = \varepsilon_d^2/[\Delta^2 + (\kappa/2)^2]$. The ratio of $n$ to the intracavity photon number on resonance gives us the following scaling factor:
\begin{equation}
\nu_2(P_{\textrm{in}}) = \frac{(\kappa/2)^2}{[\Delta(P_{\textrm{in}})]^2+(\kappa/2)^2}.
\label{eq:SI_pscale2}
\end{equation}
\indent Taking into account the contributions of both the resonator frequency shift, and the finite dynamic range of our room-temperature microwave components, we find the rescaled input power
\begin{equation}
\tilde{P}_{\textrm{in}}=\nu(P_{\textrm{in}})P_{\textrm{in}},
\label{eq:SI_pscale_f}
\end{equation}
with $\nu(P_{\textrm{in}}) = \nu_1(P_{\textrm{in}})\nu_2(P_{\textrm{in}})$. The total scaling factor $\nu(P_{\textrm{in}})$ is plotted in Fig.~\ref{fig:figSI_PhotonNumberCal}(c). The resulting rescaled input power $\tilde{P}_{\textrm{in}}$ is simply related to the photon number through a scalar multiplication factor, which we calibrate by fitting the qubit frequency shift. We use Eq.~(1) of the main text and define the effective Josephson energy as 
\begin{equation}
\tilde{E}_J = E_J J_0\left (p\varphi_{zpf} (\tilde{P}_{\textrm{in}}/{P_0})^{1/2} \right),
\label{eq:SI_EJ_eff}
\end{equation}
where $P_0$, the power associated with a single photon, is the only fit parameter. For the measurement in Fig.~3 of the main text, we find $P_0 = 41.2$~nW/photon. The linear relation between $n$ and $\tilde{P}_{\textrm{in}}$ is plotted in Fig.~\ref{fig:figSI_PhotonNumberCal}(d).

\section{Transmon Anharmonicity in the Inverted Regime}
\begin{figure}[h]
\includegraphics{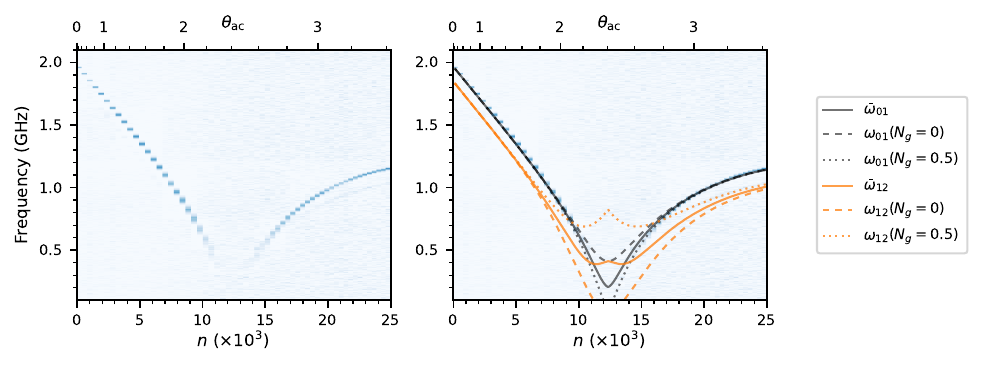}
\caption{\textbf{Transmon spectroscopy through collapse and inversion.} (Left) Frequency shift of the qubit transition measured by two-tone spectroscopy as a function of resonator occupation. The color represents the digitizer signal in arbitrary units. (Right) Same as left, with the charge averaged $\bar{\omega}_{01}$, $\bar{\omega}_{12}$ and charge-localized $\omega_{01}(N_g)$, $\omega_{12}(N_g)$ transition frequencies.}  
\label{fig:check_cpb}
\end{figure}

In this section we present the simulation results for the transmon anharmonicity in the inverted regime. On Fig.~\ref{fig:check_cpb} we show the charge-averaged spectrum of the transmon measured during the collapse and inversion. As detailed in the main text, we fitted this spectrum with the 0-1 transition frequency obtained from the diagonalization of Hamiltonian Eq.~(1) for offset charge $N_g\in\{0,0.5\}$. Here we overlay not only the 0-1 but also the 1-2 transition, for $N_g\in\{0,0.5\}$ and charge-averaged.

We identify the faint feature below the 0-1 transition in the inverted regime to be the 1-2 transition. It is detuned by the anharmonicity $\alpha=\omega_{01}-\omega_{12}$ which, near the collapse, depends strongly on the offset-charge (see the orange dashed lines ending up on either side of the black dashed lines at the collapse). As for the charge-averaged value of $|\alpha|$, when coming from the inverted regime it goes down at first (see the solid orange line coming closer to the black line). This trend can be observed in the spectroscopic signal, and constitutes a signature of the dynamical Cooper-pair box regime.

\newpage
\section{Characterization of Drive-Induced State Transitions}  
\label{sec:SI_DUST}

\begin{figure}[b]
\includegraphics{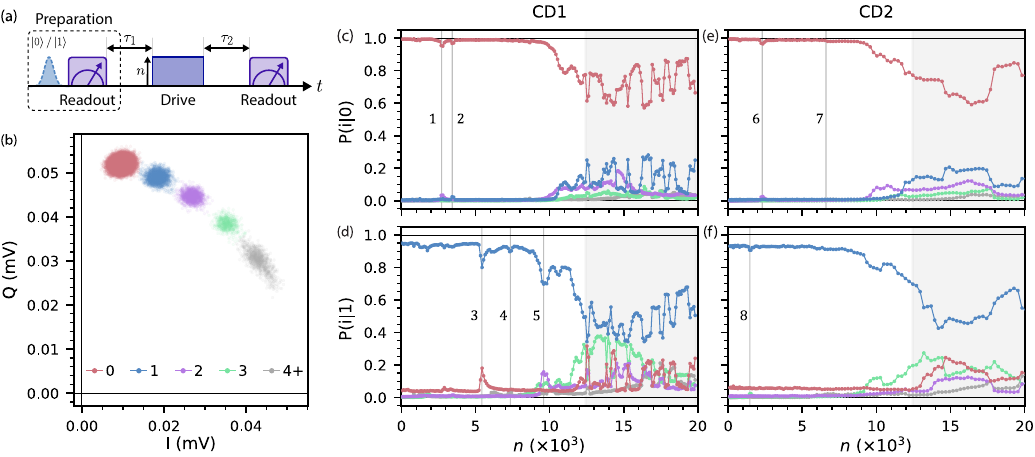}
\caption{\label{fig:figSI_DUST} \textbf{Drive-Induced State Transitions over Two Cooldowns.} (a) Pulse sequence used to measure transmon occupation probabilities after application of a 752 ns rectangular drive pulse of varying power. The frequency of the drive is set to 8.05363~GHz in cooldown (CD) 1 and 8.0539~GHz in CD2. The shape of the readout pulses, and depletion times $\tau_1$ and $\tau_2$ differ between CD1 and CD2. (b) Example of an IQ histogram for preparation in $|0\rangle$ for a single drive power $n$. We can distinguish the states up to and including $|3\rangle$. All shots above are labeled as $|4+\rangle$. For CD1, we plot the occupation probabilities as a function of $n$ for preparation in $|0\rangle$ and $|1\rangle$ in (c) and (d), respectively. All data in the main text, and supporting data in the supplementary information, is taken during CD1. (e) and (f) show the occupation probabilities measured in CD2, a previous cooldown. The most prominent drive-induced state transition events in (c)-(f) are indicated by vertical lines and labeled from 1 to 8. The grey-shaded area indicates the region beyond the collapse of the potential.} 
\end{figure}

In Fig.~4(b) of the main text, several drive-induced transitions from the excited state were identified at specific photon numbers. Here, we investigate these drive-induced state transitions in more detail to give a potential explanation for their origin. We compare two cooldowns (CD), CD1 and CD2. All data of the main text and supplementary information is taken in CD1. CD2 corresponds to a previous cooldown.\\
\indent For both cooldowns, we measure transmon occupation probabilities as a function of resonator photon number using the pulse sequence in Fig.~\ref{fig:figSI_DUST}(a). After preparation of the transmon in $|0\rangle$ or $|1\rangle$, state transitions are stimulated by a 752 ns long rectangular resonator drive pulse of varying power $n$. After a wait time of $\tau_2$, which passively depletes the resonator, the resulting transmon state is probed with a readout pulse that can distinguish states up to and including $|3\rangle$, see Fig.~\ref{fig:figSI_DUST}(b). State discrimination is performed by projecting the IQ data onto the angle in phase space and fitting Gaussians to the one-dimensional histograms of each state. The intersections of the Gaussian fits determine the discrimination thresholds. For states above $|3\rangle$, this fitting procedure fails, and all shots above are labeled as $|4+\rangle$. There are minimal differences between the pulse sequences used in CD1 and CD2. In CD1, three-segment readout pulses are used for readout, allowing a shorter depletion time $\tau_1 = 5/\kappa$, while $\tau_2 = 10/\kappa$. In CD2, the readout pulses are rectangular and $\tau_1 = \tau_2 = 10/\kappa$. Despite these small differences, the stimulation drive is the same in both cooldowns, allowing a fair comparison of the measured drive-induced excitations.\\
\indent 
The occupation probabilities as a function of photon number are plotted in Fig.~\ref{fig:figSI_DUST}(c)-(f). The general trend we observe is that the probability remains close to 1 before decreasing significantly as the photon number goes beyond the potential collapse point, indicated by the grey-shaded area. At the collapse, the energy spectrum turns gapless at $N_g=0.0$ and $N_g=0.5$. As the stimulation pulse drives the system through the collapse and back, Landau-Zener transitions \cite{Grifoni1998,Ivakhnenko2023} between the gapless Floquet modes lead to a decrease of the measured occupation probabilities. The exact transition probability depends on the offset charge, which sets the energy gaps. Surprisingly, when preparing in $|0\rangle$ ($|1\rangle$) we also observe significant population in $|2\rangle$ ($|3\rangle$), while the energy gap between these states should be significant for all $N_g$. An explanation could be given by successive Landau-Zener transitions, e.g. during ring-up and ring-down of the resonator, if the charge parity changes during the stimulation drive. Indeed, in ref. \cite{Chowdhury2026} they show theoretically that the parity lifetime can become on the order of 100~ns after crossing the collapse point in a flux-driven transmon. Overall offset-charge fluctuations are suspected to be the origin of the oscillations in $P(i|0)$ and $P(i|1)$ in CD1 beyond the collapse point. The probabilities measured in CD2 show less abrupt changes, which could be explained by the `dwell' time at a single photon number in the experiment. In CD2, a smaller range of the charge dispersion is sampled, as we only take $2.5\times10^4$ shots at each drive power, compared to $4\times10^4$ shots for CD1. \\
\indent Before the collapse, small dips in the occupation probabilities are observed in both CD1 and CD2. We will investigate these in more detail, starting with CD1. For preparation in $|0\rangle$, we identify transitions $|0\rangle\rightarrow|2\rangle$ (1) and $|0\rangle\rightarrow|1\rangle$ (2). When preparing in $|1\rangle$, the three most prominent transitions are $|1\rangle\rightarrow|0\rangle$ (3), $|1\rangle\rightarrow|4+\rangle$ (4), and $|1\rangle\rightarrow |2\rangle$ (5). All of these, except transition 4, involve at most 2 transmon excitations. Recalling that the drive frequency is $\omega_d/2\pi = 8.05363$ GHz and the bare transmon qubit frequency is $\omega_{01}/2\pi=1.972$ GHz, it becomes clear that these transitions cannot be mediated by intrinsic multi-excitation resonances, i.e. through the absorption of one or multiple drive photons. Following ref. \cite{Dai2026}, this leaves two other options for these transitions: resonant exchange with extrinsic spurious modes, such as two-level systems (TLSs), and inelastic scattering involving extrinsic spurious modes, e.g. electromagnetic modes of the environment. Transition 3 can most likely be categorized as a resonant exchange with a TLS. This transition is only observed when the transmon is prepared in $|1\rangle$, but not in $|0\rangle$, which would be consistent with energy exchange with a TLS in its ground state. Furthermore, this transition is not visible in CD2, in line with the common observation of TLS rearrangement through thermal cycling \cite{deGraaf2020}. The available data does not allow us to make a more detailed identification of the origins of transitions 1,2 and 5. Studying their dispersion with drive frequency could shed light on their nature \cite{Dai2026}.\\
\indent Next, comparing CD2 to CD1, it can be seen that the number of drive-induced state transitions in CD2 is substantially lower. Furthermore, the transitions are weaker, i.e., the amount of population transferred is less. This variability between cooldowns likely results from changes in the extrinsic mode spectrum. We identify from $|0\rangle$ the transitions $|0\rangle \rightarrow |2\rangle$ (6) and $|0\rangle \rightarrow |4+\rangle$ (7). From $|1\rangle$, the most prominent one is $|1\rangle \rightarrow |3\rangle$ (8). For transitions 6 and 8, the same considerations apply as for transitions 1,3-5 in CD1. Transitions 1 and 6 occur at similar photon numbers, potentially indicating inelastic scattering with a spurious mode that is not affected by thermal cycling, such as a mode in the electromagnetic environment of the chip. Transitions 4 and 7 are the only potential intrinsic multi-excitation resonances, which could be mediated by the absorption of at least one drive photon. It is, however, impossible to exactly determine which transmon states are involved.\\
\indent In conclusion, the above analysis shows that the majority of the drive-induced state transitions observed in our system are of extrinsic origin. This strengthens our confidence in the robustness of the circuit implementation to intrinsic multi-excitation resonances, and shows that the dominant limitations come from interactions with extrinsic spurious modes. Most importantly, even though the observed transitions result in a small decrease of the QND fidelity, the incurred population loss after their diabatic traversal is so weak that they do not spoil the observation of the collapse and inversion of the Josephson potential. 

\bibliography{bibliography}